\documentclass[usenatbib, twocolumn, a4paper,nofootinbib]{aastex63}

\usepackage[utf8]{inputenc}  
\usepackage[T1]{fontenc}
\usepackage{graphicx,psfrag}
\usepackage{mathrsfs}
\usepackage{amsmath,amsfonts,amssymb,pifont,gensymb}
\usepackage{multirow,enumerate}
\usepackage{comment,hyperref}
\usepackage{color}
\usepackage{acronym}
\usepackage{xspace}
\usepackage[normalem]{ulem}
\usepackage{mathtools}
\usepackage{subfigure}
\usepackage{makecell}
\usepackage{changepage}
\usepackage{appendix}
\usepackage{lipsum}
\usepackage{natbib}

\usepackage{color}

\newcommand{\boldtheta}{\boldsymbol{\theta}}
\newcommand{\boldLambda}{\mathbf{\Lambda}}
\newcommand{\Mgal}{\mathcal{M}_{\mu, t}}
\newcommand{\Rgal}{\mathcal{H}_{t}}
\newcommand{\Ugal}{\mathcal{W}_{\mu, t}}
\newcommand{\golum}{\textsc{GOLUM}}
\newcommand{\clu}{\mathcal{C}_U^L}
\newcommand{\cmgal}{\mathcal{C}_{\mathcal{M}_{\mu, t}}}
\newcommand{\cugal}{\mathcal{C}_{\mathcal{W}_{\mu, t}}}
\newcommand{\crgal}{\mathcal{C}_{\mathcal{H}}}
\newcommand{\cmgaldt}{\mathcal{C}_{\mathcal{M}_{t}}}
\newcommand{\cugaldt}{\mathcal{C}_{\mathcal{W}_{t}}}
\newcommand{\cluModA}{\mathcal{C}_{\rm{Mod A}}}
\newcommand{\cluModB}{\mathcal{C}_{\rm{Mod B}}}
\newcommand{\cluModC}{\mathcal{C}_{\rm{Mod C}}}
\newcommand{\cluModD}{\mathcal{C}_{\rm{Mod D}}}

\begin{document}

\title{Ordering the confusion: A study of the impact of lens models on gravitational-wave  lensing detection capabilities}

\author{Justin Janquart$^{*}$}
\email{$^{*}$j.janquart@uu.nl}
\affiliation{Nikhef – National Institute for Subatomic Physics, Science Park, 1098 XG Amsterdam, The Netherlands }
\affiliation{Institute for Gravitational and Subatomic Physics (GRASP), Department of Physics, Utrecht University, Princetonplein 1, 3584 CC Utrecht, The Netherlands}

\author{Anupreeta More}
\affiliation{The Inter-University Centre for Astronomy and Astrophysics (IUCAA), Post Bag 4, Ganeshkhind, Pune 411007, India}
\affiliation{Kavli Institute for the Physics and Mathematics of the Universe (IPMU), 5-1-5 Kashiwanoha, Kashiwa-shi, Chiba 277-8582, Japan}

\author{Chris Van Den Broeck}
\affiliation{Nikhef – National Institute for Subatomic Physics, Science Park, 1098 XG Amsterdam, The Netherlands }
\affiliation{Institute for Gravitational and Subatomic Physics (GRASP), Department of Physics, Utrecht University, Princetonplein 1, 3584 CC Utrecht, The Netherlands}

\begin{abstract}
\noindent
When traveling from their source to the observer, gravitational waves can get deflected by massive objects along their travel path. When the lens is massive enough and the source aligns closely with the line-of-sight to the lens, the wave undergoes strong lensing, leading to several images with the same frequency evolution. These images are separated in time, magnified, and can undergo an overall phase shift. Searches for strongly-lensed gravitational waves are already ongoing. In essence, such searches look for events originating from the same sky location and having the same detector frame parameters. However, the agreement between these quantities can also happen by chance, when the uncertainty on the parameters is such that they overlap, leading to an important confusion background. To reduce the overlap between background and foreground, one can include lensing models, enforcing the lensing parameters to also be consistent with our expectations. In principle, such models should decrease the confusion background. However, when doing realistic searches, one does not know which model is the correct one to be used. The use of an incorrect model could lead to the non-detection of genuinely lensed events. In this work, we investigate under realistic conditions how one can identify lensed events in the unlensed background. We focus on the impact of the addition of a model for the lens density profile and investigate the effect of potential errors in the modelling. We show that it is extremely difficult to identify  lensing confidently without the addition of a lens model. We also show that slight variations in the profile of the lens model are tolerable but a model with incorrect assumption about the underlying lens population causes significant errors compared to assuming no lens model at all. We also suggest some strategies to improve the confidence in the detection of strong lensing for gravitational waves. 
\end{abstract}

\section{Introduction}
Compact binary coalescences (CBCs) originate from the encounter of two massive objects (black holes or neutron stars) circling each other before merging, distorting the fabric of space-time. Over the last years, the Advanced LIGO~\citep{TheLIGOScientific:2014jea} and the Advance Virgo~\citep{TheVirgo:2014hva} have detected 90 such events~\citep{LIGOScientific:2021djp}. In addition, the KAGRA detector~\citep{Somiya:2011np, Aso:2013eba,Akutsu:2018axf,Akutsu:2020his} has joined the network of ground-based detectors. A fifth detector is now also being built in India~\citep{LigoIndia}. The increased sensitivity has enabled an increasing number of gravitational wave (GW) detections, as well as more and more accurate tests of general relativity~\citep{LIGOScientific:2021sio}, more accurate cosmological studies~\citep{GWTC3cosmology}, and more accurate merger rate reconstructions and representation of the mass functions for the massive objects~\citep{LIGOScientific:2021psn}. As the sensitivity increases and the network of detectors extends, the observation of new physical effects will become possible. One such phenomenon could be GW lensing. 

If a massive object (such as a galaxy, a galaxy cluster, or other compact objects) is situated along the travel path of a GW, it can deflect the wave, leading to gravitational lensing~\citep{Ohanian1974, Degushi1986, Wang:1996as, Nakamura1998, Takahashi:2003ix, Dai:2017huk, Ezquiaga:2020gdt}. Depending on the nature of the lens, the effect produced on the wave can be different. A fraction of the lensed events will undergo strong lensing~\citep{Nakamura1998, Takahashi:2003ix}, where the GW is split into several distinct images appearing in the data as repeated events~\citep{Wang:1996as, Haris:2018vmn}. This can happen, for example, for galaxy lenses~\citep{Dai:2016igl, Ng:2017yiu, Li:2018prc, Oguri:2018muv} or for galaxy cluster lenses~\citep{Smith:2017mqu, Smith:2018gle, Smith:2019dis, Robertson:2020mfh, Ryczanowski:2020mlt}.  For strong lensing, the size of the lens is typically much larger than the GW wavelength, leading to the so-call geometric optics limit, where the frequency evolution of the wave is unchanged from one image to the other. The images have the same frequency evolution and come from the same sky location. However, some of the parameters are biased by the lensing effect. The images are magnified, leading to a bias for the observed luminosity distance~\citep{Dai:2017huk, Ng:2017yiu, Pang:2020qow}. In addition, the GW can undergo an overall phase shift, which depends on the relative position of the source and the lens~\citep{Ezquiaga:2020gdt}. Finally, the different images have a time delay, leading to images arriving from minutes to months apart from each other. For smaller lenses, where the geometrical optic limits are not respected, frequency-dependent effects occur, giving rise to microlensing~\citep{Takahashi:2003ix, Cao:2014oaa, Lai:2018rto, Christian:2018vsi, Singh:2018csp, Hannuksela:2019kle, Meena:2019ate, Pagano:2020rwj, Cheung:2020okf, Kim:2020xkm}. Since massive lenses such as galaxies and clusters of galaxies are not made of a single object but also contain smaller objects, a strongly-lensed GW event can also be micro-lensed~\citep{Seo:2021ucd}.

If detected, strong lensing could open the door to new scientific opportunities. The detection of a quadruply-lensed event combined with the identification of its electromagnetic (EM) counterpart could lead to the possibility to identify the host galaxy for merging black holes~\citep{Smith:2017mqu,Hannuksela:2020xor,Wempe:2022zlk}. In addition, the combination of the two information channels (GW and EM) could enable high precision cosmography measurements~\citep{Sereno:2011ty, Liao:2017ioi, Cao:2019kgn, Li:2019rns, Hannuksela:2020xor}. The comparison between the GW lensing time delays and the EM time delays enables us to probe the speed of gravity~\citep{Baker:2016reh, Fan:2016swi}. Even if the EM counterpart cannot be identified, GW lensing opens new avenues. Indeed, when multiple images are detected, it is effectively the same as seeing the same event with an extended network of detectors. This can be used to probe the entire GW polarization content and look for potential additional polarizations predicted by alternative gravity theories~\citep{Goyal:2020bkm}. The detection of a lensed event with higher-order modes could also open the door to enhanced tests of general relativity and lead to even better sky localization capabilities~\citep{Janquart:2021nus}.

When searching for strong lensing, the main idea is to look for pairs of events that have matching detector-frame parameters and sky location~\cite{Goyal:2021hxv, Wong:2021lxf}. This can be done by comparing the likelihood of the lensed and the unlensed hypotheses, meaning comparing the likelihood of the two events to be lensed, or for similarities to have come about by chance. To do this, several parameter estimation-based tools exist. First, there is the posterior overlap method~\citep{Haris:2018vmn}. This focuses on a subset of parameters, makes a Gaussian kernel density estimation (KDE) fit, and then compares the fits to see if the posteriors are consistent with each other as expected for lensing. To further discriminate between lensed and unlensed events, it also folds in the time delay information, verifying the compatibility of the observed time delay with the expected distributions for the lensed and unlensed hypotheses. Another approach is to sample the full joint likelihood for the events, leading to higher precision~\citep{Liu:2020par, Lo:2021nae}. In this case, it is also possible to fold population effects into the analysis so as to formally compare the lensed and the unlensed hypotheses in the light of population models~\citep{Lo:2021nae}. A third method, equivalent to the full joint parameter estimation approach under the lensed hypothesis, consists in recasting the lensing likelihood as a conditioned likelihood. This enables to decrease the computational burden of the problem while keeping a high precision~\citep{Janquart:2021qov}. Several of these search methodologies have already been used to search for lensing in the LIGO and Virgo data~\citep{Hannuksela:2019kle, LIGOScientific:2021izm}. In addition, new methodologies to characterize the lenses at the origin of the observed lensed events are also being developed~\citep{Wright:2021cbn}.

One of the major bottlenecks faced when looking for strongly-lensed multiplets is the number of events and the risk of false alarms associated with it. Indeed, in principle, when searching for lensed events, one would need to verify all the pairs of events one can make out of the detected events.\footnote{And from there the triples, quadruples, ... In addition, one could also look for sub-threshold counteparts~\citep{Li:2019osa, McIsaac:2019use} which would increase the number of pairs to consider. Such candidates are not considered in this work.} However, at their design sensitivity, the LIGO and Virgo detectors could observe up to $\mathcal{O}(1000)$ events~\citep{Oguri:2018muv, Li:2018prc, Ng:2017yiu, Wierda:2021upe}, and one would need to analyze $\mathcal{O}(5 \times 10^5)$ pairs when looking for strong lensing. Naturally, the higher the number of events, the more likely it becomes to observe events with compatible parameters by chance~\citep{Wierda:2021upe, Caliskan:2022wbh}. 

Studies have been performed to assess the rate of lensing as well as the importance of the false-alarm probability~\citep{Mukherjee:2021qam, Wierda:2021upe, Caliskan:2022wbh}. In general, these studies show that once the observation time grows and the number of events considered increases, the false-alarm probability (FAP) when searching for strong lensing increases rapidly. In~\citep{Wierda:2021upe}, it is shown that the FAP grows as the square of the observational time when one only compares the frequency evolutions of the signals. This growth can be reduced by adding time delay information, hence by using a prior on time delays motivated by the expected values for a given lens model. On the other hand, in~\citet{Caliskan:2022wbh}, the FAP for lensing is evaluated based on matches between the masses of the two events under consideration, their sky location as well as the observed phase difference. They show that once $\mathcal{O}(1000)$ events are observed, it is nearly impossible to identify confidently strong lensing as the FAP reaches 1. These two studies focus on a standard setup with the detectors at design sensitivity. Therefore, they also neglect some effects that could make the identification of strongly lensed pairs event more difficult. Indeed, in realistic observation scenarios, detectors have down-time periods, and the event is seen by a reduced number of detectors. This will increase the width of the posterior distributions and make for a higher chance of agreement between the parameters. Moreover, the noise power spectral density (PSD) of the detectors can undergo changes over time, leading to periods with a louder background or lower background noise. This will also have an impact on the constraint we get on the parameters. 

In parallel with the characterization of the FAP, there have also been efforts to find the expected characteristics of strongly-lensed GWs~\citep{Haris:2018vmn, Wierda:2021upe, More:2021kpb}. This is often done by making a source population and a lens population and verifying the distributions of the lensing parameters (magnification ratios, time delays, and Morse factors) we observe for a given lens model. The inclusion of such results in the strong-lensing search pipelines should decrease the risk of false alarms~\citep{Haris:2018vmn, Wierda:2021upe, More:2021kpb, Caliskan:2022wbh}. Indeed, one would not require only the match between the frequency evolution of the signals but also that the way in which they are linked is consistent with our expectations for a given lens model. So, for example, two events with similarly compatible frequency evolution but with a time delay larger than what we expect for lensed signals would be penalized. Previous work~\citep{Wierda:2021upe} has shown that the FAP is indeed reduced. However, this was done in a simple context and one assumed that the correct model was known and directly applied. For a real search, one would not know which is the object at the root of the lensing phenomenon, making it impossible to know which model is the correct one. In addition, our models are also subject to errors that could impact the resulting observed distribution. Therefore, it is also important to investigate what the impact of an error in the model or the use of an erroneous model can have on our ability to detect strong lensing.

In this work we investigate how to extract lensed events out of a significant confusion background. We look at the impact of moving from analyzing the agreement between the posteriors of a subset of of parameters to the comparison of the entire frequency evolution. Then, we look at the impact of the inclusion of a lens model and what happens when there are errors in the model that is used. This paper is structured as follows. In Section~\ref{sec:StrongLensing}, we present the basics of strong lensing. In the next sections, we give an overview of lensing statistics and how they can be used in parameter estimation. In Section~\ref{sec:setup}, we present the setup of this study, and in Section~\ref{sec:Results} we present our results and discuss them. Finally, we give our conclusions and perspectives in Section~\ref{sec:Conclusions}.

\section{Strong gravitational wave lensing}
\label{sec:StrongLensing}

Strong gravitational wave lensing splits a GW into several images. Those have the same frequency evolution but can be (de)magnified, time-shifted and undergo an overall phase shift. For each image, the lensed and unlensed waveform can be related as~\citep{Dai:2017huk, Ezquiaga:2020gdt}
\begin{equation}\label{eq:LensedWF}
    \tilde{h}_L^j(f; \boldtheta, \mu_j, t_j, n_j) = \sqrt{\mu_j} \, e^{(2\pi i f t_j- \pi i n_j sgn(f))}\tilde{h}_U(f, \boldtheta) \, ,
\end{equation}
where $\tilde{h}_U(f, \boldtheta)$ is the unlensed waveform, dependent on the usual BBH parameters $\boldtheta$, and $\tilde{h}_L^j(f; \boldtheta, \mu_j, t_j, n_j)$ is the lensed waveform for the $j^{th}$ image, which depends on the lensing parameters. These parameters depend on the lens itself, as well as the source-lens configuration. The relative magnification $\mu_j$ corresponds to a focusing of the wave. Its value corresponds to the inverse of the determinant of the lensing Jacobian matrix~\citep{Schneider:1992, Haris:2018vmn, More:2021kpb}. The time delay $t_j$ of the image is due to the deflection of the wave, which takes a different geometrical path, changing the time of travel from the source to the observer~\citep{Schneider:1992, Haris:2018vmn, More:2021kpb}. Finally, the wave can also undergo an overall phase shift, called the Morse phase, translated by a discrete Morse factor $n_j$~\citep{Dai:2017huk, Ezquiaga:2020gdt}. It can take only three values: $0, 0.5$, and  $1$, and splits the images into different types. When $n_j = 0$, there is no shift and the image is of type I. This corresponds to the image passing via the minimum of the lens Fermat potential. If $n_j = 0.5$, one observes a type II image. In such a case, the wave is Hilbert-transformed. This happens when the wave passes via a saddle point of the potential. Finally, there are type III images, when the wave passes via a maximum of the potential. This leads to a sign flip of the wave. 

The Morse phase is degenerate with the dominant mode of the waveform and is therefore usually not measurable. However, when higher-order modes are present, this degeneracy is lifted, and image type identification becomes possible. If detected, this could lead to smoking-gun evidence for lensing~\citep{Wang:2021kzt, Lo:2021nae, Janquart:2021nus, Vijaykumar:2022dlp}.

Except for the Morse factor, the lensing parameters can be recast as \textit{effective} BBH parameters and they cannot be measured individually. So, the relative magnification can be included in an \textit{observed} luminosity distance $d_L^{\rm{obs}, j}$ and the time delay is included in an \textit{observed} time of coalescence $t_c^{\rm{obs}, j}$
\begin{equation}\label{eq:obsDl}
    d_L^{\rm{obs}, j} = \frac{d_L}{\sqrt{\mu_j}} \, ,
\end{equation}
\begin{equation}\label{es:obsTc}
    t_c^{\rm{obs}, j} = t_c + t_j \, ,
\end{equation}
where $d_L$ and $t_c$ are the unlensed luminosity distance and time of coalescence. 

Because of these degeneracies between lensed and unlensed parameters, when several images are present, one often links them using the measurable relative lensing parameters~\citep{Dai:2017huk}
\begin{equation}\label{eq:LinkLensedImages}
    \tilde{h}_L^{j}(\boldtheta, \mu_j, t_j, n_j) = \sqrt{\mu_{ji}} \, e^{(2\pi i f t_{ji} - \pi i n_{ji} sgn(f))} \tilde{h}_L^{i}(\boldtheta, \mu_i, t_i, n_i) \, ,
\end{equation}
where $\mu_{ji} = \mu_j / \mu_i$, $t_{ji} = t_j - t_i$, and $n_{ji} = n_j - n_i$ are the relative magnification, the relative time delay, and the relative Morse factor. When several images are detected, these parameters are measurable and they have an expected distribution for a given lens model~\citep{Haris:2018vmn, Wierda:2021upe, More:2021kpb}. This could be used to better make the distinction between the unlensed background and genuinely lensed events.

\section{Lensing statistics}
\label{sec:LensingStat}
For a given lens model, one can compute the expected distribution for the observed lensing parameters. Often, one focuses on the time delay as this is well determined in the lensing models and accurately measured in the gravitational-wave data~\citep{Haris:2018vmn, Wierda:2021upe}. However, the expected distributions for the relative magnification and the Morse factor can also be computed (even if the computation for the relative magnification can be subject to more debate based on the flux ratio anomalies observed in the EM observations, see for example~\citet{Mao2004, Xu2009, Xu2015, Macci2006, Hsueh:2017nlk}), which can lead to further constraints on the nature of the observed signals~\citep{More:2021kpb}. One can also compute the distributions for unlensed events~\citep{Haris:2018vmn, Wierda:2021upe, More:2021kpb}. For a given event pair, one can then compare the probability to get the observed values under the lensed and the unlensed scenarios, enabling one to rapidly focus on pairs that are compatible with the lens models. 

For example,~\citet{Haris:2018vmn} considers a single isothermal sphere ~\citep[SIS,][]{Witt:1990} lens and computes the time delay distribution for the lensed case, while assuming a Poissonian distribution for the time of arrivals for the unlensed case. From there, they make the following statistic
\begin{equation}\label{eq:Rgal}
    \mathcal{H}_{t} = \frac{p(t_{ji} | \mathcal{H}_L)}{p(t_{ji} | \mathcal{H}_U)} \, ,
\end{equation}
where $\mathcal{H}_L$ stands for the lensed hypothesis and $\mathcal{H}_U$ for the unlensed hypothesis. This statistic is used to further discriminate between lensed and unlensed events when they already established matching detector frame parameters and sky location using the posterior overlap method.

\citet{More:2021kpb} go a step further by using a single isothermal ellipsoid model ~\citep[SIE,][]{Koopmans2009} and computing the distributions for the time delays, relative magnification and relative Morse factor. The statistic
\begin{equation}
    \mathcal{M}_{\mu, t, n} = \frac{p(\mu_{ji}, t_{ji}, n_{ji} | \mathcal{H}_L)}{p(\mu_{ji}, t_{ji}, n_{ji} | \mathcal{H}_U)} \, ,
\end{equation}
is a more powerful statistic as it imposes more constraints on the observed lensing parameters. The expected relative Morse factor is dependent on which two images of the lensed multiplet are observed. Therefore, it is more difficult to use and we focus on the $\Mgal$ statistic obtained for the relative magnification and the time delay only in this work.

Similarly,~\citet{Wierda:2021upe} uses an SIE model for the lenses, except that shear is also accounted for. This addition leads to a broadening of the expected relative magnification distribution.  As for the $\Mgal$ statistic, the distributions for time delay and relative magnification can be used as classification statistics or used to further discriminate between lensed and unlensed events when the agreement between the parameters has already been established. The statistic obtained using this catalog is denoted $\mathcal{W}_{\mu, t}$ in this work.

\begin{figure*}
    \centering
    \includegraphics[keepaspectratio, width=0.49\textwidth]{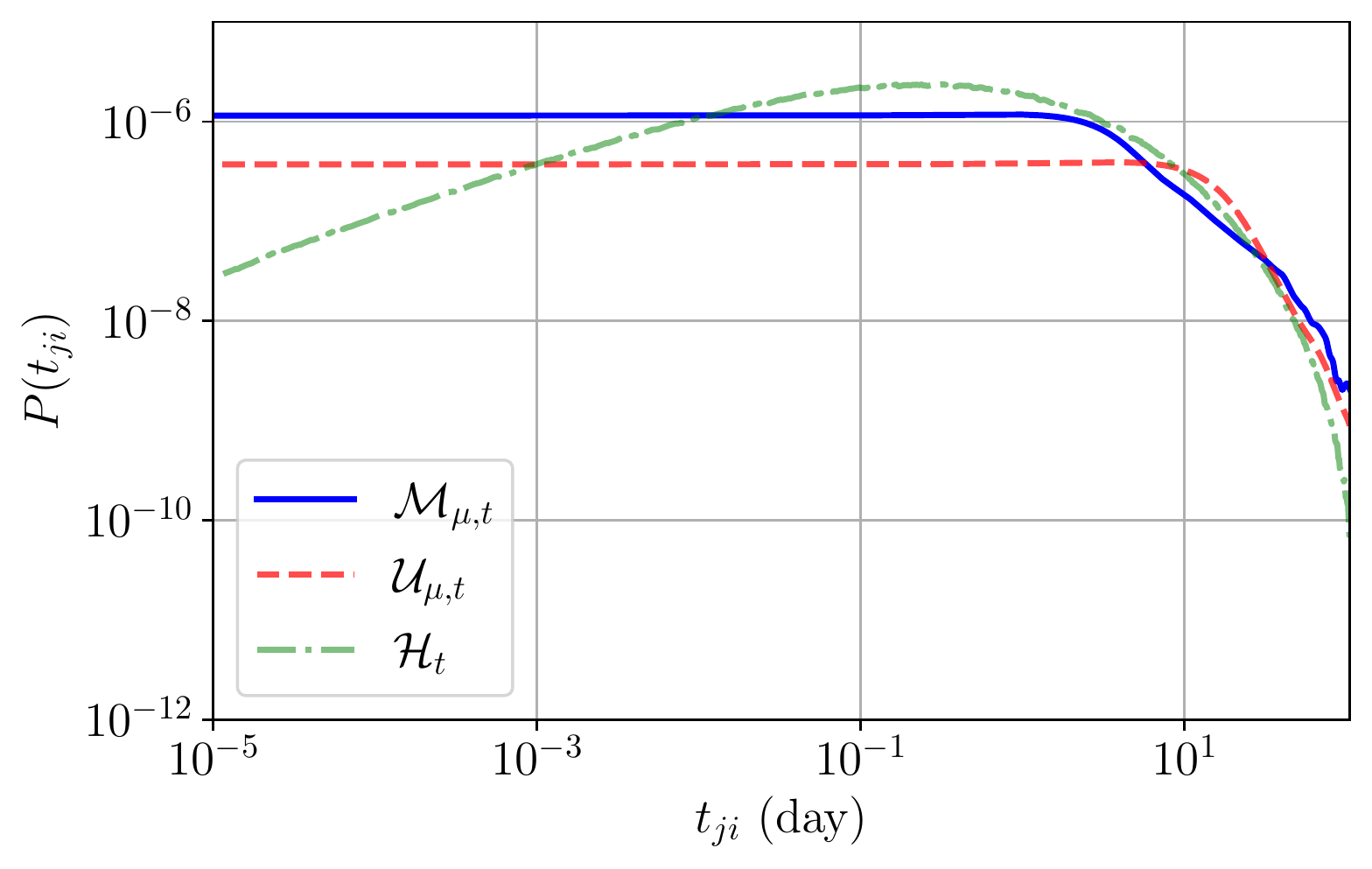}
    \includegraphics[keepaspectratio, width=0.49\textwidth]{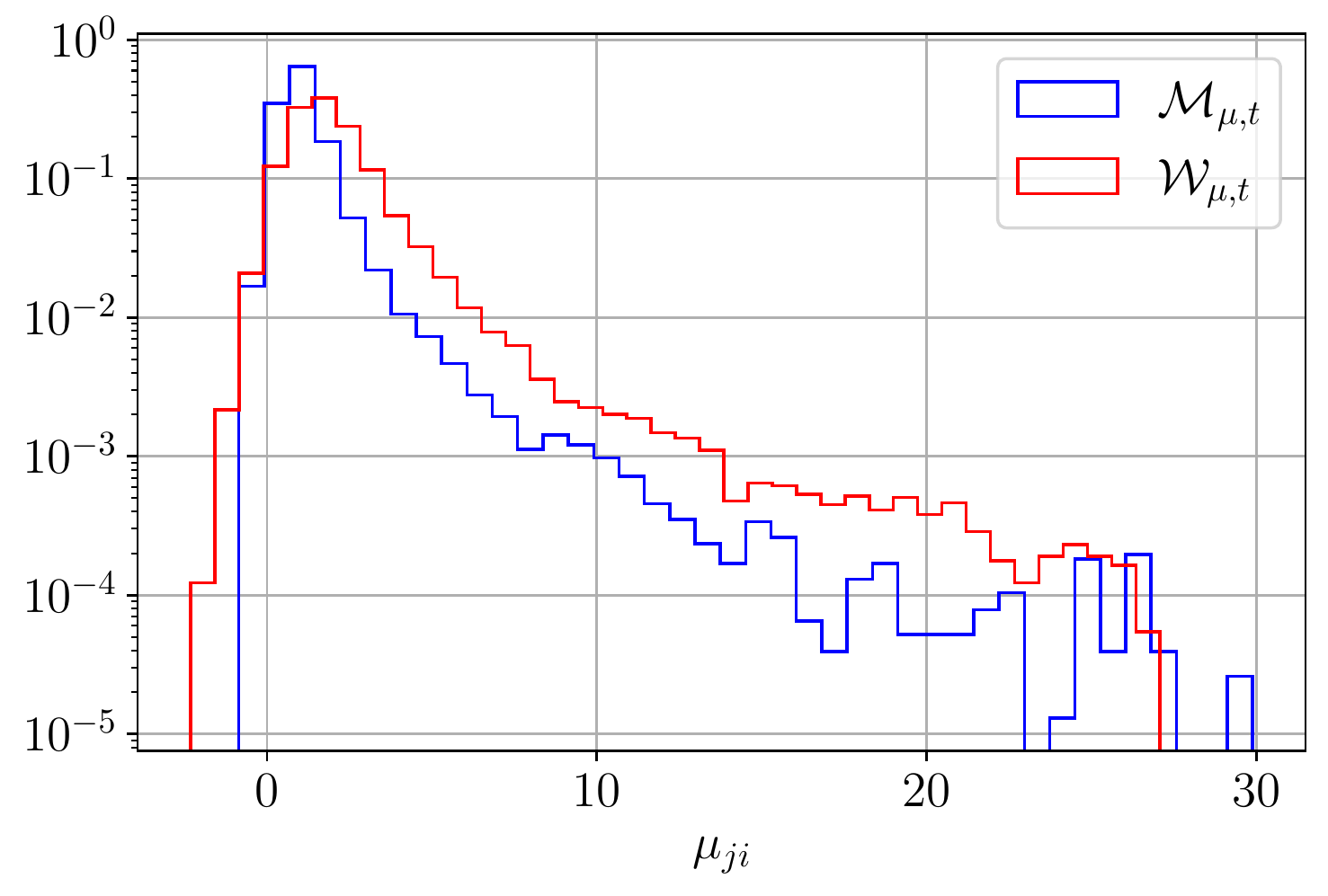}
    \caption{\textit{Left:} Time delay probabilities for the different catalogs. One sees that the SIS model is peaked at lower values before dropping fast when going to higher values. The SIE-based models have a wider probability distribution. \textit{Right:} Relative magnification distributions for the two SIE-based models. The two distributions are consistent and one sees that the addition of shear leads to a broadening of the distribution.}
    \label{fig:probMusDtsMgal}
\end{figure*}

A comparison of the distributions obtained for these different models can be seen in Fig.~\ref{fig:probMusDtsMgal}

On their own, the lensing statistics can be used to narrow down possible lensing candidates that should be followed up by more extensive pipelines. However, coupled with the lensing parameter estimation pipelines, they could significantly decrease the false alarm probability in lensing searches~\citep{Wierda:2021upe, Caliskan:2022wbh}. In particular,~\citet{Caliskan:2022wbh} shows that it would be nearly impossible to identify lensing without including the appropriate lensing statistics.

\section{Bayesian analysis for strong gravitational wave lensing}
\label{sec:BayesAnalysisForGW}
Under a given hypothesis $\mathcal{H}_{I}$, the observed data stream in the detector for an event $i$ is
\begin{equation}\label{eq:datastreams}
    d(t) = n(t) + h^i_{I}(\vartheta_{I, i}) \, ,
\end{equation}
where $I = U$ for the unlensed hypothesis and $I = L$ for the lensed hypothesis, and $\vartheta_{I, i}$ is the set of parameters needed to describe the event $i$ entirely under the hypothesis $I$. 

To compare two hypotheses, one uses the ratio of evidence under each of these hypotheses. Neglecting selection effects, the lensing evidence for two images is~\citep{Liu:2020par, Lo:2021nae, Janquart:2021qov}
\begin{align}\label{eq:LensingEvidence}
p(d_1, d_2 | \mathcal{H}_L) =& \int p(d_1 | \boldtheta, \boldLambda_1)p(d_2 | \boldtheta, \boldLambda_2) \\ \nonumber
& \times p(\boldtheta, \boldLambda_1, \boldLambda_2){\rm d}\boldtheta {\rm d}\boldLambda_1{\rm d}\boldLambda_2 \, ,
\end{align}
where $\boldtheta$ are the usual BBH parameters, and $\boldLambda_i$ are the lensing parameters for the $i^{th}$ image, $p(d_i | \boldtheta, \boldLambda_i)$ is the likelihood for image $i$, and $p(\boldtheta, \boldLambda_1, \boldLambda_2)$ are the joint priors. 

For the unlensed hypothesis, the evidence is
\begin{equation}\label{es:UnlensedEvidence}
    p(d_1, d_2 | \mathcal{H}_U) = \int p(d_1 | \boldtheta_1) p(d_2 | \boldtheta_2) p(\boldtheta_1) p(\boldtheta_2){\rm d}\boldtheta_1 {\rm d}\boldtheta_2 \, ,
\end{equation}
where $p(\boldtheta_i$) are the individual priors on the parameters for image $i$.

To decide whether an event is lensed or not, these evidences are compared in the \textit{coherence ratio}
\begin{equation}\label{eq:Clu}
    \mathcal{C}_U^L = \frac{p(d_1, d_2 | \mathcal{H}_L)}{p(d_1, d_2 | \mathcal{H}_U)} \, ,
\end{equation}
which quantifies the similarity between the signals. This does not include any selection effects which would include information about the population effects in the lensed and unlensed hypotheses; see~\citet{Lo:2021nae} for more information on selection effects.

Generally, the coherence ratio is computed using non-informative priors on the lensing parameters (e.g. uniform in time delay and magnification). This choice is made to keep generic searches unbiased towards a particular model of the lens density profile. Indeed, including  distributions predicted by a given model leads to results valid only for that particular model primarily. Even though typical lens galaxies can be well-fit with an SIE model, groups-to-cluster scale lenses can have complicated mass distributions and assuming an incorrect lens model could bias the detectability of lensed events in the data. However, one can still use the results obtained from a model-agnostic run and convert them into model-specific results. 

If $\mathcal{Z}^{M}_{\mathcal{H}_I}$ is the evidence for a given model $M$ under the hypothesis $\mathcal{H}_I$, and $\mathcal{Z}^{R}_{\mathcal{H}_I}$ is the evidence obtained from the run for the same hypothesis, then (see Appendix~\ref{sec:ChangeEvidenceDerivation} for a more detailed derivation)
\begin{equation}\label{eq:EvidenceConversion}
    \mathcal{Z}^{M}_{\mathcal{H}_I} = \bigg\langle \frac{p(\vartheta_I | M, \mathcal{H}_I)}{p(\vartheta_I | R, \mathcal{H}_I)} \bigg\rangle_{p(\vartheta_I | D, R, \mathcal{H}_I)} \mathcal{Z}^R_{\mathcal{H}_I} \,.
\end{equation}
In this expression, $p(\vartheta_I | M, \mathcal{H}_I)$ and $p(\vartheta_I | R, \mathcal{H}_I)$ are the probability to observe the parameters in the model and in the run for a given hypothesis, while $p(\vartheta_I | D, R, \mathcal{H}_I)$ is the posterior distribution obtained from the model-agnostic run for the data $D$. 

In practice, one does not solve the integral over the ratio of probabilities but uses the samples obtained from the runs to compute the weights for each set of samples and then take the average (hence performing a Monte Carlo integration). So, 
\begin{equation}\label{eq:NumericalEvidenceConversion}
    \mathcal{Z}^{M}_{\mathcal{H}_I} = \frac{1}{N} \bigg( \sum_{i = 0}^{i = N} W^{M}_{R}(\vartheta_I^i, \mathcal{H}_I) \bigg) \mathcal{Z}^{R}_{\mathcal{H}_I} \, ,
\end{equation}
where $N$ is the total number of samples obtained from the run, and
\begin{equation}\label{eq:WeightsNum}
    W^{M}_{R}(\vartheta_I^i, \mathcal{H}_I) = \frac{p(\vartheta_I^i | M, \mathcal{H}_I)}{p(\vartheta_I^i | R, \mathcal{H}_I)}
\end{equation}
is the ratio of probabilities between the model and the run for a set of parameters $i$. Here, the ${\{\vartheta_i}\}$ samples are drawn from the run posterior distibution $p(\vartheta_I | D, R, \mathcal{H}_I)$.

The model-dependent coherence ratio is then obtained by taking the ratio of the evidence for the lensed and the unlensed hypotheses, hence
\begin{equation}\label{eq:ModelDepentClu}
    \mathcal{C}_U^L\bigg|_{Model} = \frac{\mathcal{Z}^{M}_{\mathcal{H}_L}}{\mathcal{Z}^{M}_{\mathcal{H}_U}} \, .
\end{equation}

In the end, since the reweighing process is faster than the parameter estimation run, this approach enables one to adapt the results for different models without increasing significantly the computational burden. In addition, if the initial coherence ratio is low, one already knows that the event is not lensed since the parameters should match regardless of the lens model and the model-dependent part of the analysis is not needed. Therefore, a good strategy would be to first carry out the parameter estimation for all of the events and then apply the reweighing to account for the effect of the various lens models for the events with a high coherence ratio\footnote{In this work, we do the reweighing exercises for all of the events, even those with a low coherence ratio as we want to see how the background and foreground change when the lens models are included.}.

\section{Injections and setup of the study}
\label{sec:setup}

\subsection{BBH population}

In this work, we study the impact of the lens model included in the coherence ratio computation on our ability to differentiate between lensed and unlensed events. Since the fraction of strongly lensed events is relatively low, $\mathcal{O}(10^{-3})$~\citep[e.g.,][]{Xu:2021bfn, Wierda:2021upe}, we focus on making an extensive unlensed background with a few lensed events on top. Therefore, we generate 100 unlensed BBH mergers. Their masses are sampled from the PowerLaw + Peak distribution \citep{LIGOScientific:2021psn}. The spins and redshifts are sampled from the ones observed by the LIGO-Virgo-KAGRA (LVK) collaboration~\citep{LIGOScientific:2021psn}. The sky location is sampled from a uniform distribution over the sky, the inclination is uniform in cosine, the phase and polarization are uniform in their domain. We take the time of arrival for the unlensed events to be uniform in a year. For each event, we draw randomly from one of the following cases - the event is observed by i) the two LIGO detectors ii) one of the LIGO detectors and the Virgo detector or iii) by the three detectors jointly. It is important to vary the number of detectors since fewer detector lead to larger uncertainty on some of the critical parameters, such as the sky location. In turn, this leads to more compatibility between the posteriors and higher coherence ratios. For each set of parameters drawn from the distribution, we take the PSD to be that of one of the events in GWTC-2.1~\citep{LIGOScientific:2021usb, GWTC2_dataRelease} or GWTC-3~\citep{LIGOScientific:2021djp, GWTC3_dataRelease}, and generate colored Gaussian noise from the PSD. We then inject the GW strain into the noise. This leads to a realistic scenario for detections where the number of detectors and the noise are different from one event to the other. The change in the observation conditions between events is important as the differences in noise and number of detectors change the accuracy we have from one event to the other, impacting the observed detection statistic.

From these 100 unlensed events, we make 1500 unlensed pairs. In addition, we add $50$ lensed event pairs. The masses, spins, and apparent luminosity distances for the first image are drawn from the same distributions. We then generate the second image by drawing the relative magnification and the time delay from the $\mathcal{M}_{\mu, t}$ parameter catalog~\citep{More:2021kpb}. From here on, we take these distributions to be the true lensing parameter distribution. 

We analyze the different events under the unlensed hypothesis using \textsc{BILBY}~\citep{Ashton:2018jfp}, and analyze the events under the lensed hypothesis using the \textsc{GOLUM} framework~\citep{Janquart:2021qov} which provides fast and accurate joint parameter estimation for strong lensing. The two parameter estimation pipelines are used with the \textsc{DYNESTY} sampler~\citep{Speagle:2020}.

\subsection{Population analyses}

Using our extensive background, we perform different analyses to better understand the process of identifying the lensed events in an unlensed background. For each event pair, we perform a posterior overlap analysis~\citep{Haris:2018vmn}\footnote{The consistency between posteriors is computed here for the component masses, the sky location, the spin amplitudes and tilt angles, and the binary's inclination, similarly to the approach followed in~\citet{Hannuksela:2019kle, LIGOScientific:2021izm}.} as well as a joint parameter estimation run~\citep{Janquart:2021qov}. The objective here is to look into the gain one has when including more parameters when comparing the two signals and the difference between the use of the posteriors only and the use of the strains.

Using the coherence ratios obtained from the joint parameter estimation run, we reweight them for several models using the procedure explained in Sec.~\ref{sec:BayesAnalysisForGW}. We use data from three different catalogs for the lensed models: the $\Rgal$ time-delay distribution~\citep{Haris:2018vmn}, the $\Mgal$ time-delay and relative magnification distribution~\citep{More:2021kpb}, and the $\Ugal$ time-delay and relative magnification distribution~\citep{Wierda:2021upe}\footnote{We used the code base from~\citet{Wierda:2021upe} but adapted the detector networks and their sensitivity to match our situation.}. For $\Mgal$ and $\Ugal$, we do the analysis once with the 2 lensing parameters included, and once with only the time delay. This enables us to probe the impact due to the addition of the relative magnification. 

We also introduce four artificial models which represent various observation scenarios. We denote these models, A, B, C, and D. Model A is constructed as a fake galaxy-cluster lens catalog, where we focus on larger relative magnifications and time delays. The two lensing parameters follow a scaled beta distribution. For the relative magnification, the distribution peaks at 10 and has a minimum and a maximum value of 2 and 30, respectively. For the time delay distribution, the peak is at 3 months and the minimum and maximum are 3 days and 1 year, respectively. Model B uses the same relative magnification distribution as the $\Mgal$ catalog but has a different time delay distribution. We take it to be a Gaussian peaking at 4 months with a standard deviation of 1.5 months. This example illustrates the impact of a mismodeling of one of the two parameters. The last two models (C and D) resemble the $\Mgal$ model as they focus on the same region of parameters space. However, model C has loose bounds, with $\mu_{ji} \in [0.02, 32]$ and $t_{ji} \in [30\,{\rm s}, 400\, {\rm days}]$, while model D has tighter bounds, with $\mu_{ji} \in [0.5, 3]$ and $t_{ji} \in [2\, {\rm hr}, 60\, {\rm day}]$. These two models represent what would happen if one uses tight or loose bounds on the lensing parameters to be more conservative or to detect more events, respectively. 

For each of these models, the probability density in the $(\mu_{ji}, t_{ji})$-plane is obtained by sampling from the distributions for the individual parameters and performing a KDE reconstruction. The consequence of this is mainly to smoothen the edges of the distributions. A summary of the various lens models used in this work is given in Table~\ref{tab:RecapModels}.

\begin{table*}[t!]
    \centering
    \begin{tabular}{c l}
    \hline
    \hline
    Model &  Description  of the model \\
    \hline
    \hline
    $\Mgal$ & SIE-based model for relative magnification and time delays described in~\citet{More:2021kpb} \\
    $\mathcal{M}_{t}$ & Same as $\Mgal$ but  where we only consider the time delay distribution \\
    $\Ugal$ & SIE + shear based model for the relative magnification and time delays described in~\citet{Wierda:2021upe} \\
    $\mathcal{W}_{t}$ & Same as $\Ugal$ but where we only consider the time delay distribution \\
    $\Rgal$ & SIS-based model for the time delay described in~\citet{Haris:2018vmn}\footnote{This model can also be defined based on an SIE but we do not do this here as the goal is to have another lens model.} \\
    Model A & Toy model representing galaxy-cluster lenses with scaled beta distributions \\
    & for relative magnification (peak at 10 with minimum of 2 and maximum of 30) and \\
    & for time-delay (peak at 3 months, minimum of 3 days and maximum of one year). \\
    Model B & Toy model with same $\mu_{ji}$ distribution as $\Mgal$ but with a shifted time delay ($\mathcal{G}(4\, \rm{months}, 1.5\,\rm{months})$).\\
    Model C & Toy model based on $\Mgal$ but using broader bounds on the lensing parameters,\\
    & with $\mu_{ji} \in [0.02, 32]$ and $t_{ji} \in [30\, {\rm s}, 400\, \rm{days}]$. \\
    Model D & Toy model based on $\Mgal$ but using tighter bounds on the lensing parameters, \\ & with $\mu_{ji} \in [0.5, 3]$ and $t_{ji} \in [2\, \rm{hr}, 60\, \rm{day}]$. \\
    \hline
    \end{tabular}
    \caption{Summary of the different lensing models used in this work.}
    \label{tab:RecapModels}
\end{table*}

For the $\mu_{ji}$ and $t_{ji}$ distributions of the unlensed events, we use the distributions given in~\citet{More:2021kpb} for the unlensed events for all of the models except for $\Rgal$. These distributions correspond to a census of magnification (i.e. distance) ratios and time delays for the unlensed pairs of BBH population and depend on the specific assumptions of the BBH population. 
For the $\Rgal$ scenario, we use the same approach as in~\citet{Haris:2018vmn}, where the unlensed time delay is modeled as a Poissonian process. For the unlensed cases in ~\citet{More:2021kpb} and in~\citet{Haris:2018vmn}, the probability density is higher for longer time delays when compared to the lensed scenario (see e.g. Fig.~2 in~\cite{More:2021kpb} and Fig.~2 in~\cite{Haris:2018vmn} for a representation). 

\subsection{Determining lensed candidates}
To determine whether events are lensed or not, we need to use some threshold on the detection statistics. One way of doing this is to use a fixed threshold on the statistic. For example, one could claim an event to be a lensed candidate as soon as $\ln{\mathcal{C}} > 2$, where $\mathcal{C}$ is any detection statistic. This is similar to the approach considered in~\citet{jeffreys_2003}. However, this is a generic approach and does not account for the characteristics of the data we are considering. Therefore, in this work, we take an approach similar to the one used in~\citet{Caliskan:2022wbh} with the false-alarm probability (FAP) given by
\begin{equation}\label{eq:FAPdef}
\rm{FAP} = \frac{\# \rm{Unlensed} > \rm{X}}{\# \rm{All \, Unlensed}} \, .
\end{equation}
Here the numerator is the number of unlensed above X, a threshold defined based on the observations in the lensed scenario, and the denominator is the total number of unlensed events. 

In this work, X is chosen to be the fifth percentile of the detection statistic for the lensed foreground. Using this method, we are able to fold in the effect of the models on both the lensed and the unlensed population. For example, if the model favors the unlensed events and disfavors the lensed ones, its impact on the statistics is such that their values increase for the unlensed events. On the other hand, they decrease for the lensed events, leading to a smaller value of the fifth percrentile. Therefore, $\# \rm{Unlensed} > X$ increases and the FAP becomes higher. 

In this work, other information is also used to characterize the performance of the detection statistic for a given model. The receiver operator characteristics (ROC) curve represent the ability of the model to differentiate between lensed and unlensed pairs. It represents the efficiency versus the false positive probability (FPP). The efficiency is defined as the fraction of lensed events having their detection statistic higher than a given value, while the FPP is the number of unlensed events with their detection statistic larger than the given value. So, one wants that highest possible efficiency for an FPP that is as low as possible. 

Another way to represent the performance is to use the complementary cumulative distribution function (CCDF) of the unlensed background with the cumulative density function (CDF) of the lensed foreground. Ideally, one wants the CCDF to drop as fast as possible, while the CDF for the lensed foreground should be significant at values as high as possible. The overlap between those two curves will represent the confusion region, where the value of the detection statistic is such that it can correspond to lensed and unlensed events. The smaller this region, the easier it is to distinguish between lensed and unlensed events.

\section{Results}
\label{sec:Results}
\subsection{From posterior overlap to joint parameter estimation}

First, we verify how the change from posterior overlap to joint parameter estimation modifies the distribution of the corresponding detection statistic. For that, we compute the overlap between the parameters for all the events in our catalog (lensed and unlensed) using the method from~\citet{Haris:2018vmn} as well as the coherence ratio using \textsc{GOLUM}~\citep{Janquart:2021qov}. 

\begin{figure*}
    \centering
    \includegraphics[keepaspectratio, width=0.49\textwidth]{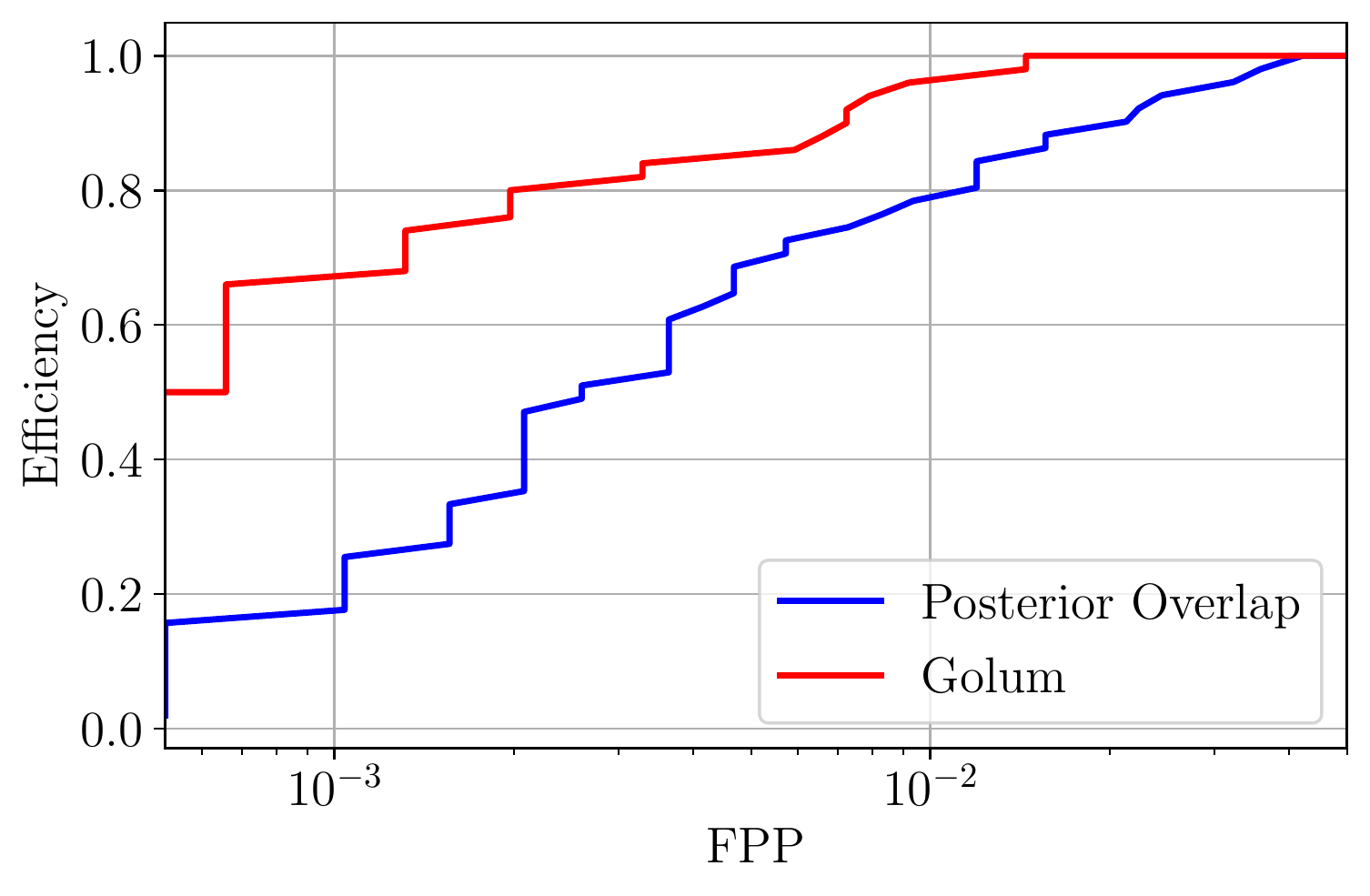}
    \includegraphics[keepaspectratio, width=0.49\textwidth]{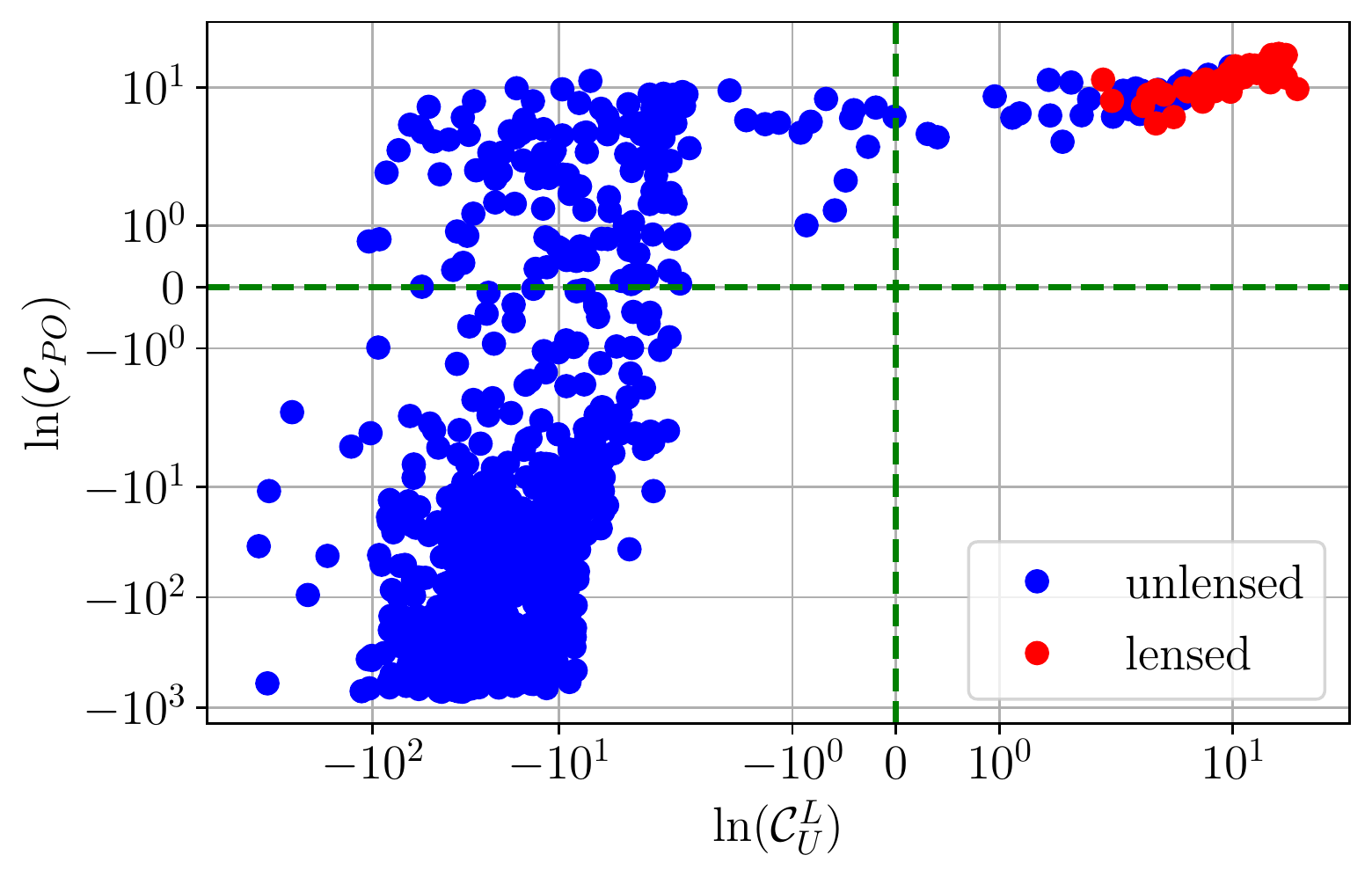}
    \caption{\textit{Left: } ROC curves for \golum~ and the posterior overlap methods. Since the curve for the joint parameter estimation tool is more to the left and reaches one faster, it means that it is better suited to determine whether events are lensed or not. \textit{Right: } Comaprison between the $\ln (\mathcal{C}_U^L)$ and the $\ln{(\mathcal{C}_{PO})}$ statistics for the lensed and unlensed events in our catalog. We see that for the lensed events, the two method seem correlated. However, there is a clear reduction in the number of high significance pairs when using joint parameter estimation.}
    \label{fig:CompaPO_Golum}
\end{figure*}

The comparison between the two is given in Fig.~\ref{fig:CompaPO_Golum}, where a ROC curve is shown as well as a scatter plot of the detection statistic for the two methods. These plots show that there is a real gain in using a framework like \textsc{GOLUM}, where one ascertains more stringently the correlation between the signals. Based on the ROC curve, we see that the FPP for a given efficiency is reduced when going from one framework to the other. This is also evident from the number of unlensed events with an $\ln{(\mathcal{C}_U^L)} > 0$ being lower for \golum~ compared to the posterior overlap. For the lensed events, the two frameworks agree relatively well. If we take the threshold for the lensing detection to be the fifth percentile of the lensed detection statistic distribution, then $\rm{FAP} = 0.85\%$ for \golum~ and $\rm{FAP} = 3\%$ for posterior overlap, showing that seeking for better correlation between the parameters of the GW signals leads to a lower risk of misidentifying an unlensed event as a lensed one.

Based on this observation, one could advocate the use of a fast joint parameter estimation tool such as \golum~ to filter out the events before doing more extensive searches. Usually, a \golum~ run would still require the full analysis of the first image, inluding the Morse factor information. This corresponds to a classical parameter estimation run and is relatively expensive. However, the image type included in the search becomes important when there is a strong higher-order mode (HOM) contribution in the observed event~\citep{Ezquiaga:2020gdt, Wang:2021kzt, Janquart:2021nus, Vijaykumar:2022dlp}. So, a preliminary strategy could be to use the posteriors obtained by the usual LVK pipelines~\citep{LIGOScientific:2021djp}, such as \textsc{BILBY}~\citep{Ashton:2018jfp}, and perform the analysis of the second image using those posteriors, by-passing the more computationally costly first image run. Under the assumption that the HOMs are weak, the distributed coherence ratio~\citep{Janquart:2021qov}
\begin{equation}\label{eq:distributedClu}
    \clu = \frac{p(d_1 | \mathcal{H}_L)}{p(d_1 | \mathcal{H}_U)}\frac{p(d_2 | d_1, \mathcal{H}_L)}{p(d_2 | \mathcal{H}_U)} 
\end{equation}
can be approximated by 
\begin{equation}\label{eq:distributedClu}
    \clu \simeq \frac{p(d_2 | d_1, \mathcal{H}_L)}{p(d_2 | \mathcal{H}_U)} 
\end{equation}
as the ratio of evidence for the first image is $\mathcal{O}(1)$ since the only difference between the two is the image type and it cannot be detected without a significant HOM contribution. 

With this method, only the run for the second image would be needed in \golum, and it would take $\mathcal{O}(30\, \rm{min})$ at most while enabling a better reduction of the background. Of course, if an event is flagged as having a significant HOM contribution, one would necessarily have to redo the joint parameter estimation completely to make sure that nothing was missed because of potential biases~\citep{Janquart:2021nus, Vijaykumar:2022dlp}\footnote{We note that if the HOM content is strong enough to significantly bias the \golum~ analysis, it would probably also bias the posterior overlap analysis, where the samples are usually taken from a standard unlensed parameter estimation run.}.

Some preliminary investigations performed on our catalog show that it is the case that most of the events are well recovered without accounting for the Morse factor in the first image. More extensive comparisons are left for future work.

\subsection{Including the correct model}
Once the catalog has been analyzed and a coherence ratio has been obtained for all the pairs, one can include the effect of lensing statistic in the final results using Eq.(\ref{eq:EvidenceConversion}). This should decrease the confusion region where the background and the foreground  overlap and hence the FAP~\citep{Haris:2018vmn, Wierda:2021upe, Caliskan:2022wbh}. In this section, we analyze what happens when the \textit{true} model is used. In our case, this means the $\Mgal$ model. We denote the detection statistic associated with the $\Mgal$ model $\mathcal{C}_{\mathcal{M}_{\mu, t}}$.

\begin{figure}
    \centering
    \includegraphics[keepaspectratio, width=0.49\textwidth]{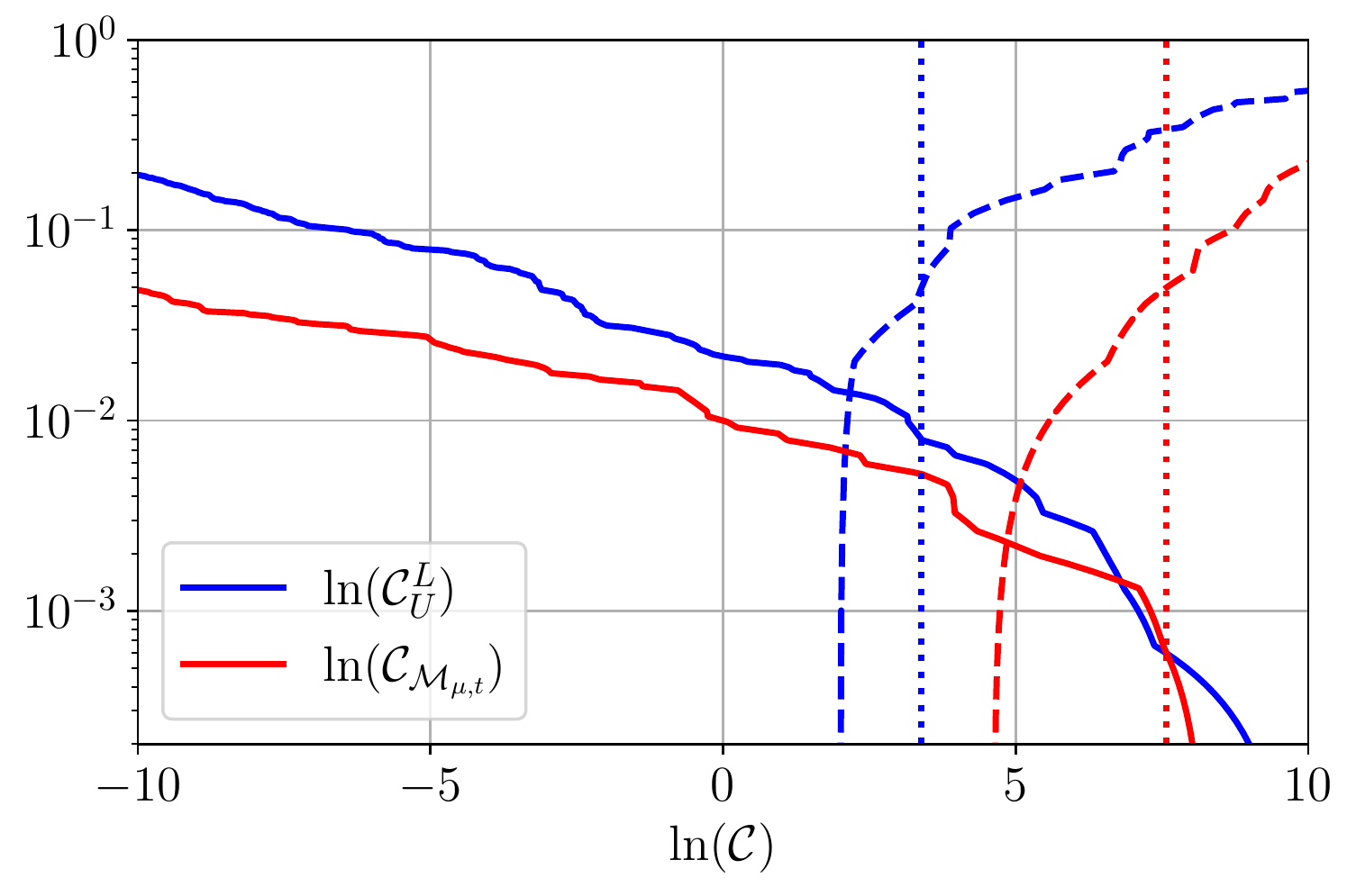}
    \caption{Comparison of the CCDF for the unlensed events (continuous lines) and the CDF for the lensed events (dashed lines) for the $\clu$ (blue)  and for the $\mathcal{C}_{\mathcal{M}_{\mu, t}}$ (red) statistics. The dotted lines are the values of the statistic for the $5^{th}$ percentile for the lensed foreground. The vertical dotted lines represent the $5^{th}$ percentile of the lensed distribution, which can be seen as the threshold above which one can consider an event to be lensed. We see that the fraction of unlensed events with a statistic higher than the fifth percentile is significantly reduced when including the lensing statistics.}
    \label{fig:CCDF_CluMgal}
\end{figure}

A comparison between the background CCDF and the foreground CDF is shown in Fig.~\ref{fig:CCDF_CluMgal}. One sees that the region of overlap between the lensed and unlensed distribution is reduced when including the lensing statistics. Indeed, the crossing between the CCDF of the unlensed events and the CDF of the lensed events happens for a higher value and encompasses a smaller area. The unlensed background is decreased for the higher values of $\clu$ but the tail is not entirely pushed back. This happens because a small number of unlensed events is promoted to a higher $\cmgal$ when the $\Mgal$ information is added. Indeed, amongst the events starting with $\ln{(\clu)}$ close to zero, some have apparent relative magnifications (i.e. their distance ratios) and time delays more compatible with the lensing hypothesis than the unlensed hypothesis purely by chance. As a consequence, those are not pushed to a lower value but a slightly higher value, mimicking quite well the lensed scenario. However, such events are in the minority and there is an effective decrease in the number of unlensed events with a high significance. For instance, we go from an $\rm{FAP} = 0.8\%$ for the $\clu$ to an $\rm{FAP} = 0.07\%$ for the $\mathcal{C}_{\mathcal{M}_{\mu, t}}$ statistic. In the end, this confirms that the inclusion of the lensing statistics helps in the reduction of the FAP and makes for more confident detections. Thi is consistent with previous studies \citep{Haris:2018vmn, Wierda:2021upe, More:2021kpb, Caliskan:2022wbh}.

We note that our values seem a bit more pessimistic than those presented in~\citet{Caliskan:2022wbh} because of the following two main reasons. The first one is the number of events that we analysed in this work. Indeed, since we need to perform parameter estimation on all of the events and all of the pairs, we do not consider as many events as analysed in \citet{Caliskan:2022wbh}. However, the goals of our works are different and yet complementary.  In \citet{Caliskan:2022wbh}, the goal was to show how difficult it is to identify genuinely lensed pairs in a large number of samples and to show how the FAP evolves with the number of samples. Our goal is to study the effect of the addition of specific lens model in the identification of lensed pairs in an unlensed background. Secondly, we consider more realistic and complex observational conditions. We use PSDs coming directly from the third observation run and vary the number of detectors observing different events. This leads to worse constraints on some of the parameters and more scope for match between the unlensed events by chance. Nevertheless, both studies suggest that it is difficult to identify lensed pairs in a background of unlensed events, even if the inclusion of a lens model can help in reducing the risk of false alarms.

\subsection{Using other models}
In the previous section, we have shown that including the expected distributions for the relative magnification and the time delay in the detection statistic helps disentangle the unlensed background and the lensed foreground. However, here, we use the underlying model used to generate the lensed events. When performing real lens searches, the lens population characteristics is not known accurately (the lens properties for a galaxy-scale or a cluster-scale lens are very different~\citep{Dai:2016igl, Ng:2017yiu, Smith:2017mqu, Smith:2018gle, Smith:2019dis} ). In addition, the models for a given type of lens can also be different, for example, several types of density profiles exist for a galaxy lens, such as SIS~\citep{Witt:1990}, SIE~\citep{Koopmans2009} and SPEMD~\citep{Barkana:1998qu}.
Although some are favored by electromagnetic observations~\citep{Koopmans2009}, there is no guarantee that the assumed model is the best representation of the true lens population in the Universe, and even the best-fitting models are subject to simplifications and inherent degeneracies. For example, we know that our prediction of the relative magnification is less robust than the one for the time delay. The former can be impacted by smaller objects present in the macro-lens~\citep[e.g.,][]{Cheung:2020okf, Yeung:2021roe}, which could lead to smeared distributions or secondary peaks. Therefore, we look at what happens when we use a different lensing statistic catalog and when we use only the time delay coming from the lensing statistics. 

\subsubsection{Effect of shear}

Here, we focus on the difference in detection statistics when including shear in the model while the underlying model has no shear. Therefore, we compare the results from the $\Mgal$ and $\Ugal$ models as the two rely on an SIE model, except that the second includes shear. We note the detection statistic based on the $\Ugal$ model $\cugal$. As shown in Sec.~\ref{sec:LensingStat} and Fig.~\ref{fig:probMusDtsMgal}, the two have relatively close distributions, and the main effect of shear is to widen the relative magnification distribution. We also note that the $\Ugal$ statistic has a slightly higher probability of large time delays. 

\begin{figure}[t!]
    \centering
    \includegraphics[keepaspectratio, width=0.5\textwidth]{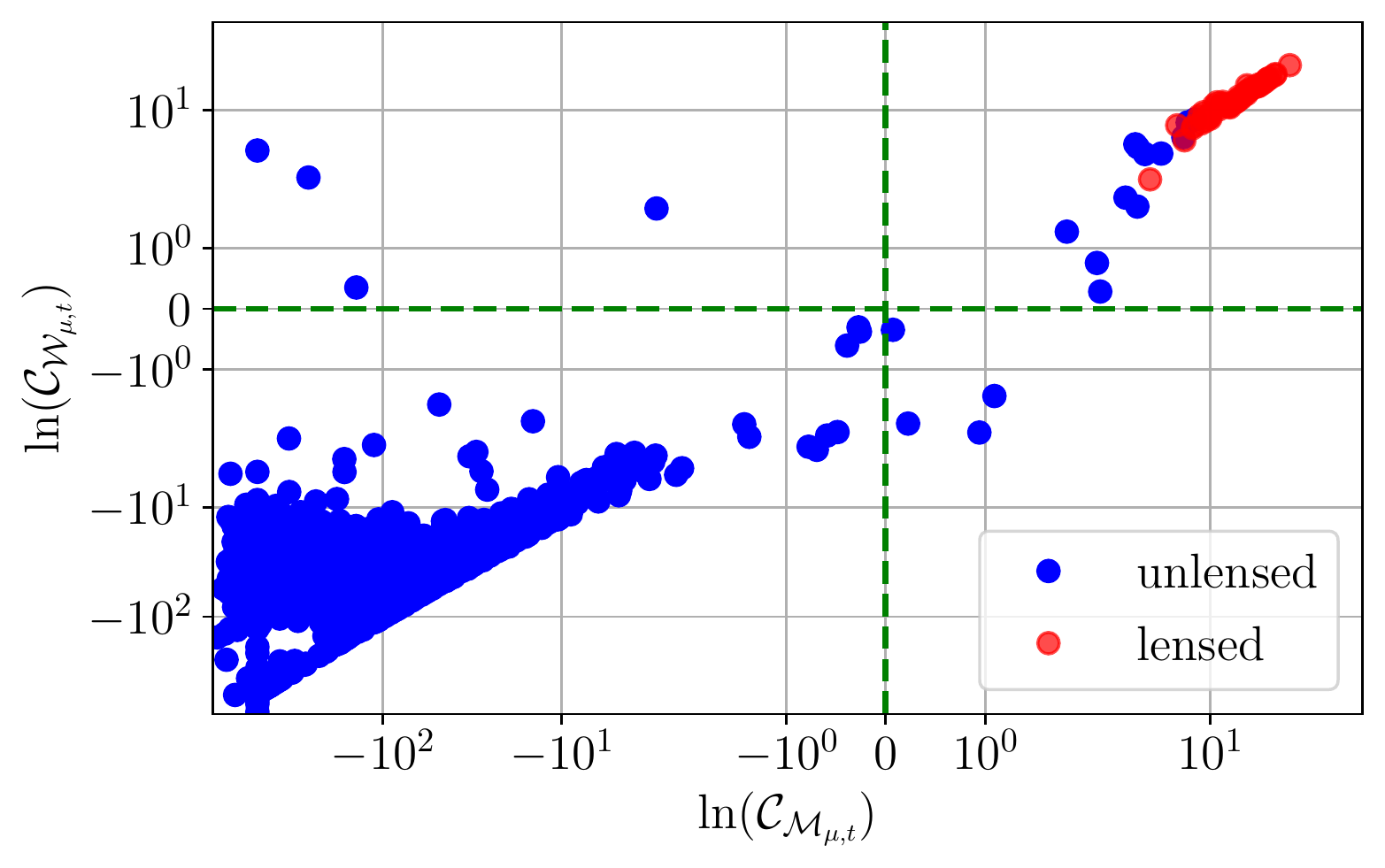}
    \caption{Comparison of the foreground and background for the $\cmgal$ and $\cugal$ detection statistics. We see that for most of the events, the two models agree reasonably well. Some events have a higher significance for one model or the other. This happens when the apparent lensing parameters are in a high probability region of the model. For the lensed events, the two models agree well, even if there is a slight decrease in significance for some events when using the $\Ugal$ distributions. Because of that, the percentile for the lensed event will decrease a bit and the FAP is slightly worse for the $\Ugal$ model, showing that there is an effect of the small disagreements between the two models.}
    \label{fig:ScatterUgalMgal}
\end{figure}

A comparison of the detection statistics for the unlensed and the lensed events for the two models is shown in Fig.~\ref{fig:ScatterUgalMgal}. One sees that the two statistics agree quite well, with a few exceptions. For the most part, the unlensed background is unchanged between the two in the sense that most of the unlensed events for one model are also categorized as unlensed for the other model. For $\cugal$, there are a few events with low $\cmgal$ that are pushed to a higher statistical values. This happens when the time delays are close to some hundred days and the relative magnification is large. Indeed, in that case, one is in the highest probability values of the model and there is a significant boost due to the lensing statistics. This is also seen for the $\cmgal$ statistic where a few events are promoted. This happens for events with short time delays and magnifications close to 1. Still, only a few of the events are significantly changed. Some other events are also a bit increased for $\cugal$, especially at a low value for $\cmgal$. This is mainly an effect of the broadening of the relative magnification, where more events become compatible with the lensed hypothesis. 

For the lensed events, we see that a few have $\cugal < \cmgal$. This is because the peak probability density is reached for different values of the time delay and the relative magnification. Still, no lensed event is entirely discarded. However, this decrease in significance for some events leads to an increase in the FAP, as the fifth percentile has a lower value and. So, more unlensed events have their value above the threshold. For the $\cugal$ statistic, the $\rm{FAP} = 0.1\%$. As a consequence, doing the same analysis with the same density profile as the underlying distribution but with slight variations in the model is still better than using no model at all. Indeed, the FAP is reduced significantly for the $\Ugal$ model when compared to the statistics for the coherence ratio.

\subsection{Using only the time delay}
As has been mentioned previously, the time delays are less sensitive to a given model compared to the relative magnification. Therefore, it can be appealing to use only the time delay to reweigh the coherence ratio.

To investigate this, we analyze the lensed and unlensed event pairs using the time delay distributions obtained for the $\Mgal$ and $\Ugal$ models and note these detection statistics $\cmgaldt$ and $\cugaldt$ respectively. In addition, to study the impact of an error on the density profile of the lens, we include the time delays obtained from the $\Rgal$ model. In this model, the lens profile is an SIS instead of an SIE, leading to a different shape of the time delay distribution (see Fig.~\ref{fig:probMusDtsMgal}).

\begin{figure*}[t!]
    \centering
    \includegraphics[keepaspectratio, width = 0.49\textwidth]{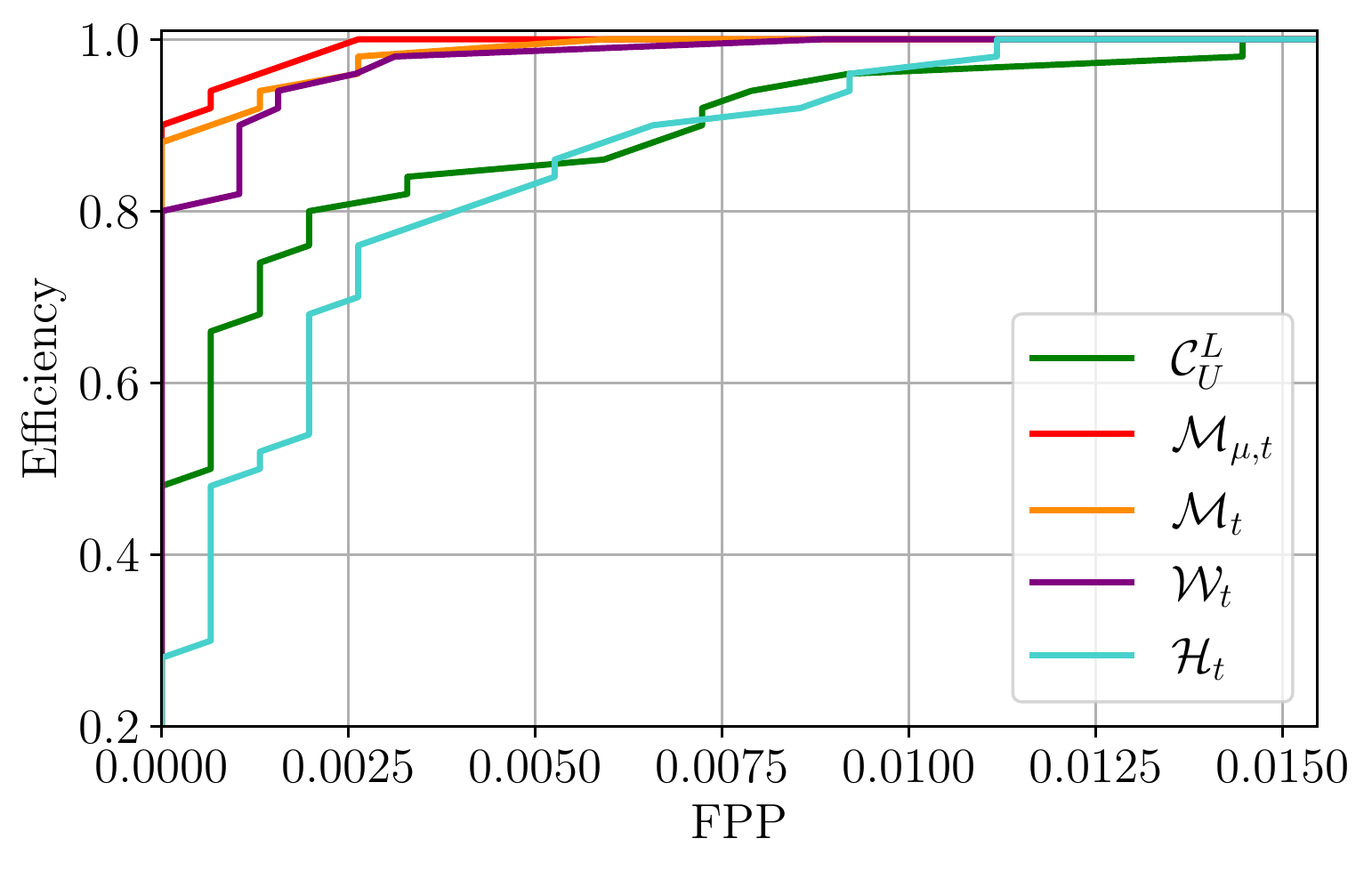}
    \includegraphics[keepaspectratio, width = 0.49\textwidth]{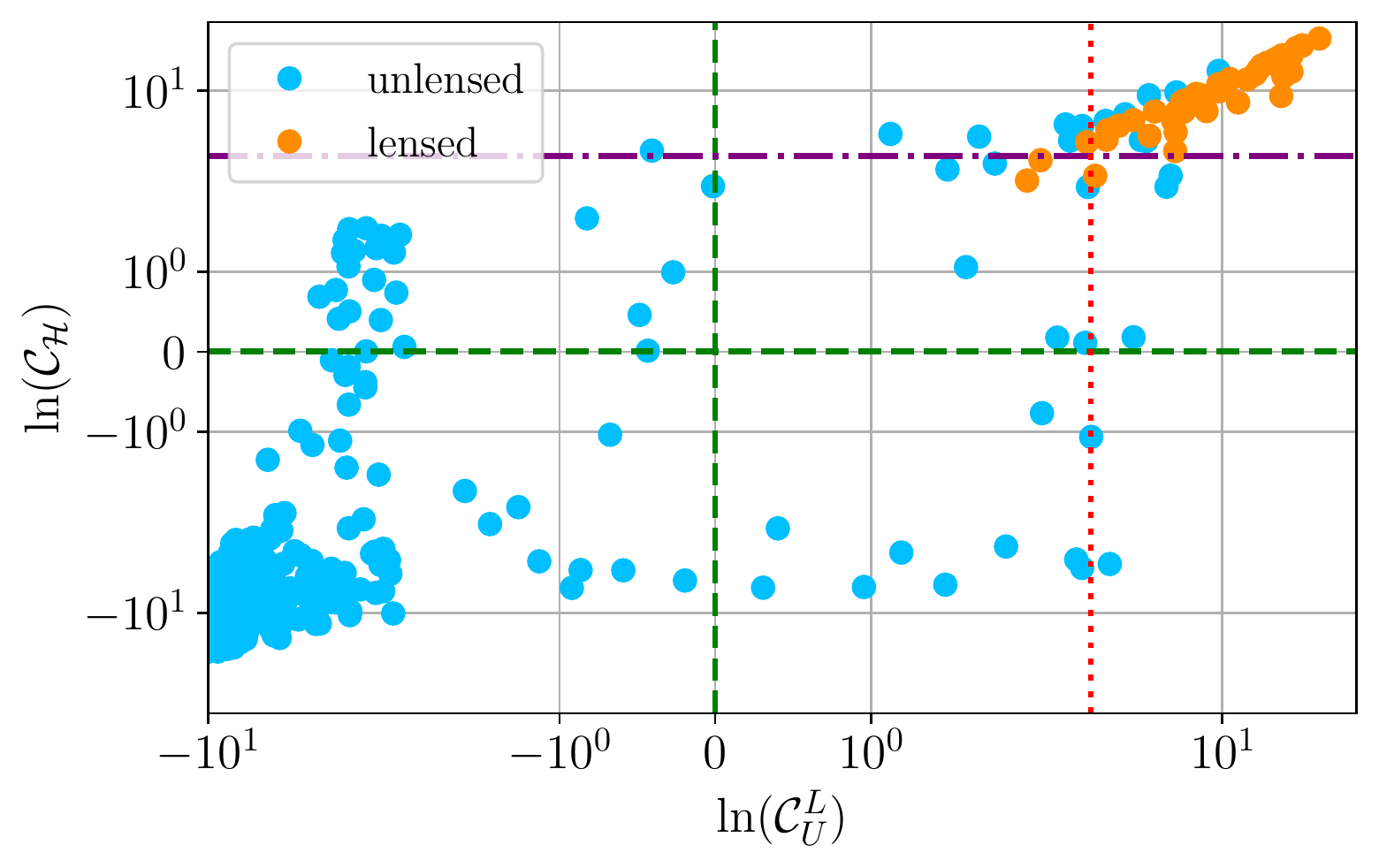}
    \caption{\emph{Left:} ROC for the different models with only the time delay included. The curves for the $\clu$ and $\Mgal$ models with relative magnification and time delay included are also added for comparison. One sees that the use of the time delay only leads to a slight loss in the performance for the search. Still, the picture remains relatively close, making the identification of lensing possible. On the other hand, the model that has an entirely different density profile for the lens ($\Rgal$) has a significantly poorer performance, performing even worse than without the inclusion of a lens model. \emph{Right: } Comparison of the detection statistics for the $\Rgal$ model and no model at all ($\clu$), with the $5^{th}$ percentile (purple dash-dotted line for $\ln{(\crgal)}$, and red dotted line for $\ln(\clu)$). One sees more unlensed events with more significant statistics for the wrong model. In addition, more unlensed events cross the $5^{th}$ percentile threshold, leading to a higher false alarm probability.}
    \label{fig:MURgal}
\end{figure*}

\begin{figure}
    \centering
    \includegraphics[keepaspectratio, width = 0.49\textwidth]{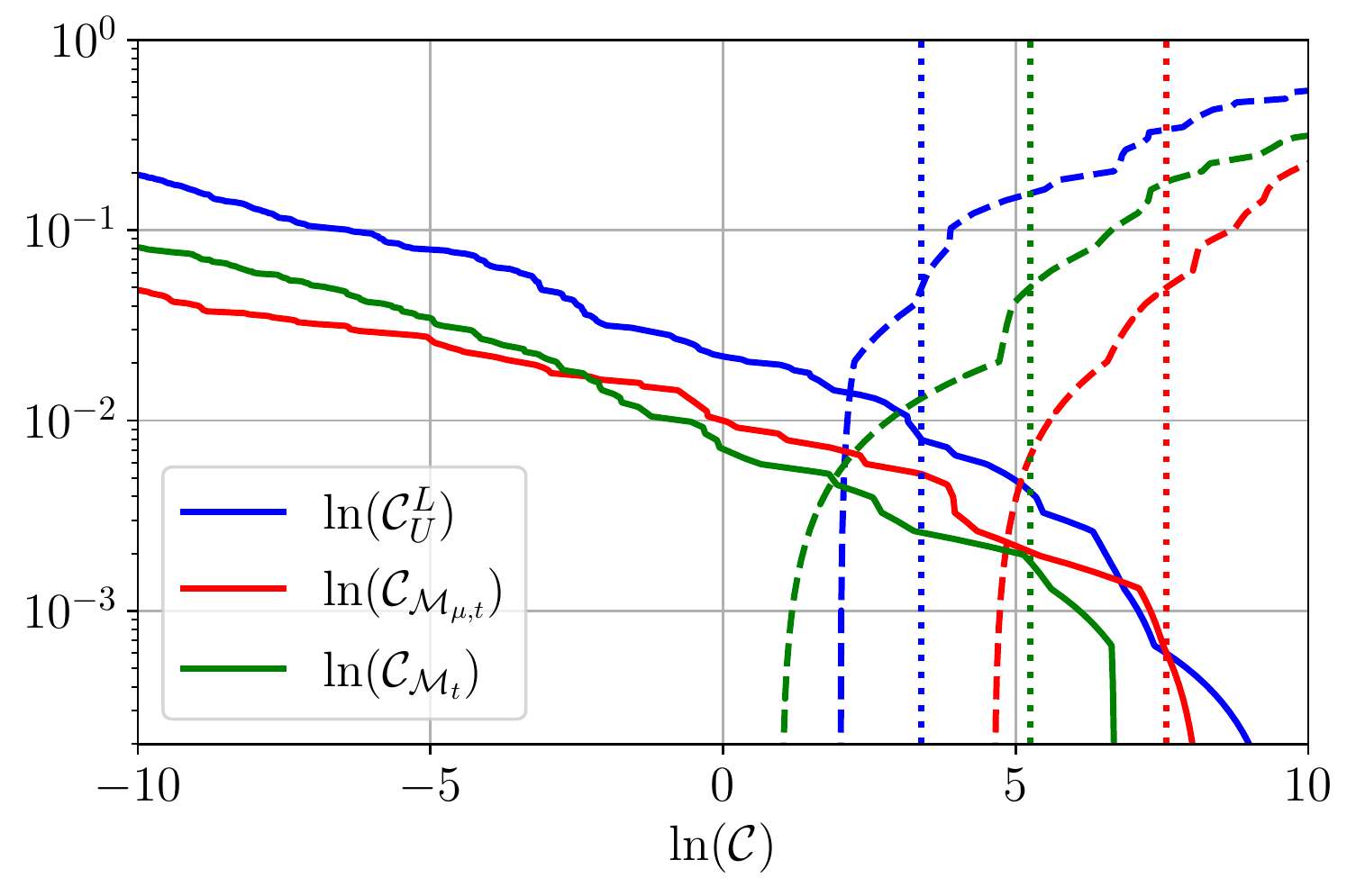}
    \caption{Comparison of the CDF and the CCDF for the background and foreground for the $\Mgal$ model with and without the relative magnification included. The dotted lines represent the $5^{th}$ percentiles of the statistics for the lensed foreground. We also represent the $\clu$ distributions for comparison. We see that the non-inclusion of the relative magnification leads to a larger confusion region. However, it sill performs much better than including no model at all.}
    \label{fig:CDF_CCDF_MgalMuDt_MgalMu_Clu}
\end{figure}

A comparison of the performances for the three models and for the $\clu$ and the $\Mgal$ model is shown in the left panel of Fig.~\ref{fig:MURgal}. One sees that there is no major difference between using the time delay for the SIE or the SIE + shear models, with a very small difference at lower FPP, which come from the lower probability for lower time delays. Still, we see that in this case, the difference between the models is smaller than for the one with the relative magnification included. The two models are slightly less efficient than the correct model including both the relative magnification and the time delay. For the $\Mgal$ model, some events have a compatible time delay but not a compatible relative magnification. Therefore, they have a lower $\cmgal$ when the reweighting is performed. Those events are not flagged here and thus increase the FPP of the background. A comparison between the $\Mgal$ model with and without magnification, and the coherence ratio in terms of CDFs and CCDFs for the background and foreground is given in Fig.~\ref{fig:CDF_CCDF_MgalMuDt_MgalMu_Clu}.  The FAPs for the $\Mgal$ and $\Ugal$ models with only the time delay included are $0.19\%$ and $0.21\%$ respectively. This is higher than the values for the same models with the relative magnification included. On the other hand, we see that the picture is much worse when including the wrong density profile with a very different shape in probability. Indeed, the curve found in the ROC plot from Fig.~\ref{fig:MURgal} is not at all comparable to the one from the other models. It is worse than for the case without model. Indeed, the two curves become comparable around an FPP of 0.005 but at lower values, the $\Rgal$-based model is worse. This can also be seen in the FAP, where, for $\crgal$, $\rm{FAP} = 0.92\%$, which is higher than for the $\clu$ statistic. 

A closer comparison between these two statistics can be seen in the right panel from Fig.~\ref{fig:MURgal}, where we represent the values for one statistic compared to the values for the other statistic. One sees that the two are not entirely correlated for higher values and that one has more unlensed pairs with a high $\crgal$ compared to $\clu$. There are also more unlensed events with a value higher than the $5^{th}$ percentile for $\crgal$.  This means that it is more difficult to make the difference between lensed and unlensed events than when no model is included. In the end, this means that including a model quite different from the real one can harm the identification of the lensed events. However, if the lensed event present in the data is a ``golden'' lensed event (with a very high $\clu$ and a relatively short time delay), the wrong statistic will still enable one to detect such an event. In this case, the detection is likely to be less confident than using the correct statistic. 

\subsection{Analyses of the toy models}
In this section, we analyze the results for the toy models. These are also important as they represent some hypothetical scenarios of interest and represent what can happen if there are major errors in the model (for example, using an entirely biased model). 

\subsubsection{Effect of important errors in the model}
Here, we look at the results for models A and B, where the two parameters are strongly biased or the time delay is biased to higher values. Such scenarios could be observed for some galaxy-cluster lenses~\citep[e.g.,][]{Smith:2017mqu, Smith:2018gle, Smith:2019dis, Robertson:2020mfh, Ryczanowski:2020mlt}. We note the statistics for models A and B are $\cluModA$ and $\cluModB$, respectively.

\begin{figure}
    \centering
    \includegraphics[keepaspectratio, width = 0.49\textwidth]{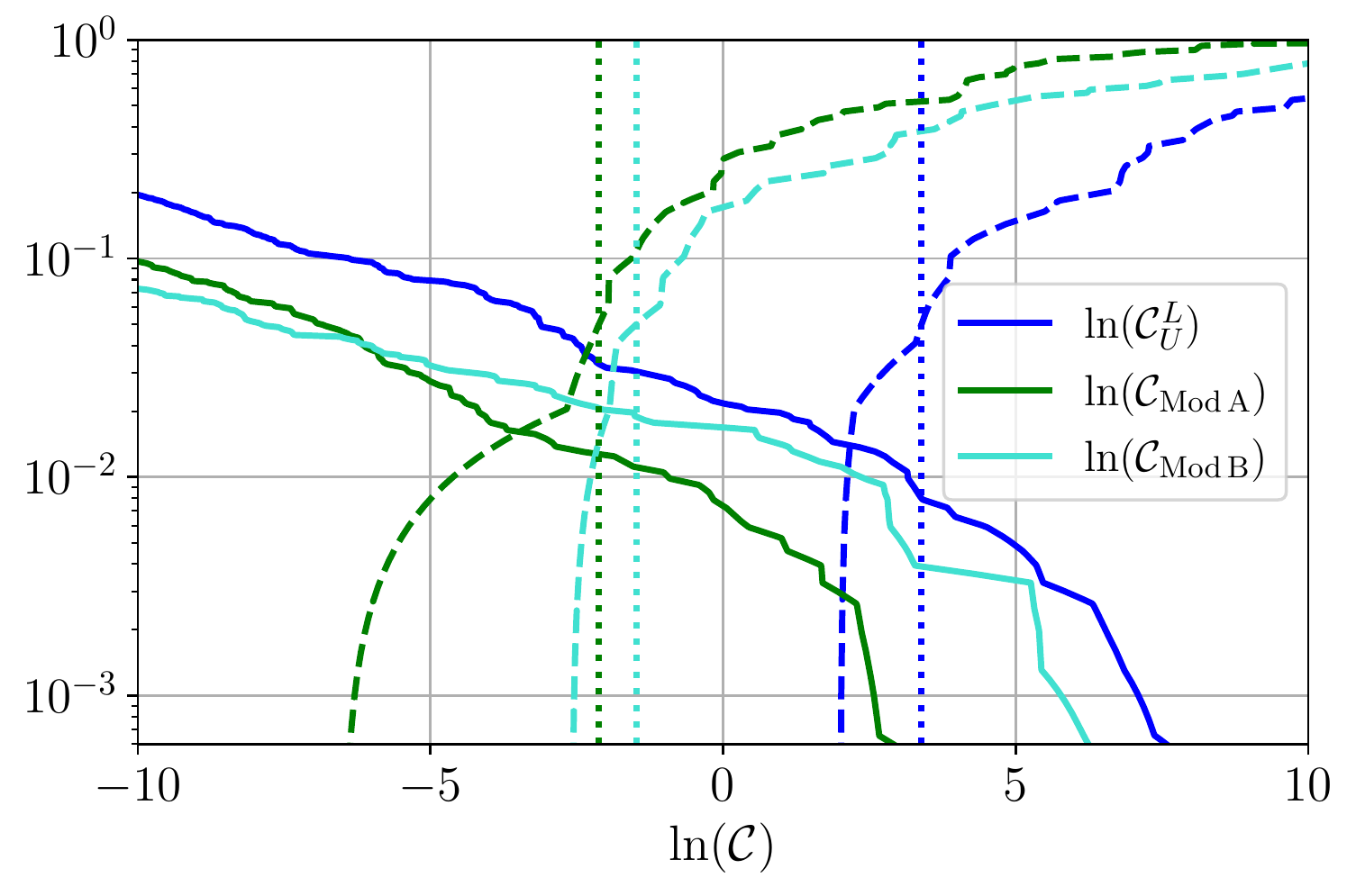}
    \caption{CDF and CCDF for the alternative models A and B, and for the coherence ratio. In this case, one sees that the models are performing worse than when no model is included. This shows that including a model that would be entirely wrong compared to the true underlying distribution would make the identification of the lensed events nearly impossible. Such a case could happen when analyzing a galaxy lensed event with a galaxy cluster lens or vice-versa.}
    \label{fig:ModAModBClu}
\end{figure}

A comparison of the CDF and CCDF for model A, model B, and the $\clu$ is given in Fig.~\ref{fig:ModAModBClu}. The two models are clearly giving worse results than when no model is used. The effect is more important than for the $\Rgal$ case. Indeed, here, not only the shapes of the distributions for the time delay are different but they are also focusing on an entirely different region of the parameter space as models A and B are configured for higher time delay values. In this case, the identification would be nearly impossible for the two models. For model B, the relative magnification has the same distribution as the underlying real distribution. Still, one sees that the resulting model is clearly worse and that having one correct parameter out of two is not enough to compensate when the other is strongly biased. Notably, one sees that some of the events get a negative $\ln{(\cluModA)}$ or a negative $\ln{(\cluModB)}$. In such a case, the identification of lensing would become extremely difficult as a significant part of the unlensed events have a higher significance than some of the lensed events. For model A, we observe an FAP of $2.3\%$, while for model B it is $2.4\%$. The higher value for the second model is explained by a higher number of unlensed events being pushed towards a larger value. Indeed, the unlensed events tend to have larger time delays, which are more compatible with the lensed distribution for this model.

The situation represented by these models could be the one faced when analyzing a strongly lensed event with one type of lens in mind (for example a galaxy cluster model) while the actual lens is something else (for example a galaxy lens). Since one does not know the true nature of the lens beforehand, it means that performing an entire analysis based only on one model could hinder the detection of a truly lensed event. In addition, in our situation, the FAP is computed when knowing the underlying true distribution. However, in reality, this is not the case. Therefore, if one is not careful and uses a model that is entirely biased, it would have a lot of high significance unlensed events, which could lead to false claims. The only way to make sure of the nature of the event would be to perform a background study to verify the significance of the candidate event. So, one would have to perform an extensive injection study and use the state-of-the-art BBH population and lens distributions. For example, one could use the BBH population given by the LVK collaboration~\citep{LIGOScientific:2021psn}, and a lens distribution taken from a catalog and compute the statistical significance of the candidate pair. This exercise would be analogous to the one presented in this work, except that the FAP would be represented by the number of unlensed pairs with a detection statistic higher than the lensed candidate under consideration\footnote{This would be one of the safest ways to ascertain the lensed nature of the event but would also be computationally extensive, as a significant number of parameter estimation runs would be required.}. 

Finally, since the time delay distribution for galaxy cluster lensed events overlaps much more with the distribution expected for unlensed events, we expect the effect of the lensing statistics to be reduced. Hence, a robust identification of a galaxy-cluster lensed event would be more difficult than for a galaxy lens \citep[see also][for similar results]{Wierda:2021upe}. 

\subsubsection{Effect of the bounds on the lensing parameters}
In this section, we focus on the alternative models C and D which have broader and tighter bounds, respectively, than the $\Mgal$ model but focus on the same region of parameter space. This could be seen as a proxy for the use of the highest and lowest bounds on the model parameters. Instead of taking a hard cut on the bounds and keeping the same probability density, we rescale it to the new bounds. Therefore, in practice, we dilute the probability density for model C and condense it for model D. We denote the detection statistic with $\cluModC$ and $\cluModD$ for the models C and D, respectively. 

\begin{figure}
    \centering
    \includegraphics[keepaspectratio, width = 0.49\textwidth]{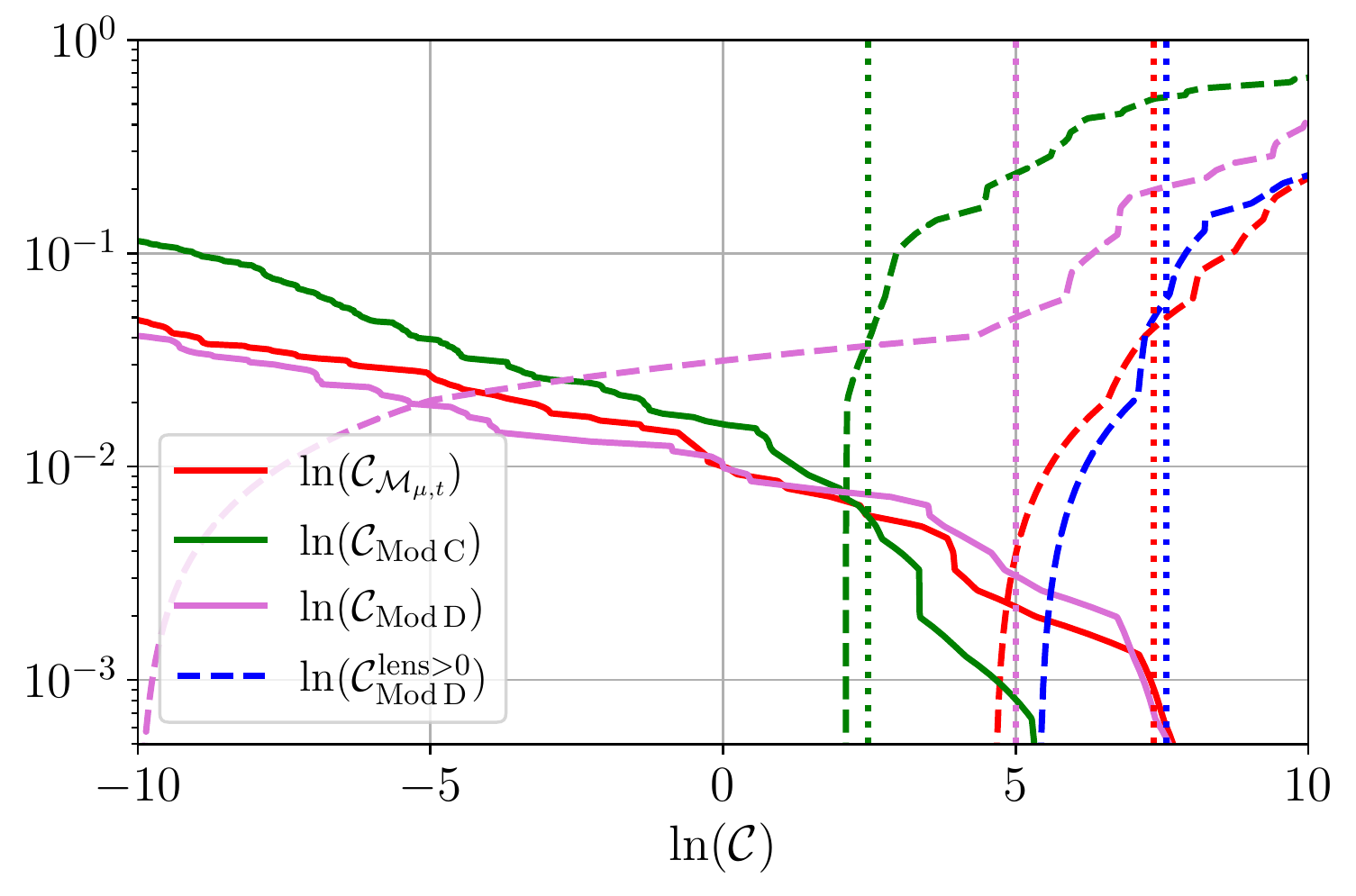}
    \caption{CCDF for the unlensed background and CDF for the lensed foreground for models C and D, and for $\Mgal$. The dashed lines represent the $5\%$ percentiles of the statistics for the lensed foreground. The pink curve for the lensed model D foreground extends to very low values because of two lensed events with their lensing parameters outside of the bounds considered for this model. When those events are neglected (or considered as background), one gets the blue curves. In this case, we see that the confusion background gets much closer to to the one obtained for the $\Mgal$ model. Model C has larger bounds and a diluted probability density in its domain of application. Therefore, we get a lower significance for the lensed events. Model C and D could represent the effect of taking the upper and lower bounds on some model parameters.}
    \label{fig:ModCModDMgal}
\end{figure}

A comparison of the performances for models C and D, and for $\Mgal$ is given in Fig.~\ref{fig:ModCModDMgal}. One sees that the change in bounds has consequences on the performance of the model. Indeed, the two alternative models have a larger confusion background, making for a harder time making the difference between lower significance lensed pairs and higher significance unlensed pairs. For model C, since the bounds on the lensing parameters are larger, it means that more of the unlensed events have lensing parameters that can be compatible with the lensed hypothesis. However, since the probability density is reduced, the unlensed events get less promoted and we get a reduction of the background for the very high values. On the other hand, the lensed events get a smaller boost from the lensing statistics and therefore reach lower values. As a consequence, we also observe an increase in the FAP, with $\rm{FAP} = 0.56\%$. This value remains lower than without including any model. However, it is more difficult to confidently identify the lensed events compared to when the exact injected model is used. For model D, the FAP increases quite a bit as well, since $\rm{FAP} = 0.83\%$. This is lower than that without including any models as for the broader bonds, but it is still higher than the FAP obtained from using the true model. This happens mainly because of a decrease in the fifth percentile for the lensed distribution. Out of the 50 lensed events, 2 have lensing parameters that have values outside of the bounds covered by model D. Therefore, they get a significant reduction in their statistic, which decreases the percentile in return. Finally, since there are still some unlensed events with compatible apparent lensing parameters, they get promoted to higher values and end up above some of the lensed events. Therefore, the background extends to values comparable to those seen for $\Mgal$. If we remove the 2 events with negative $\ln(\cluModD)$, then, the FAP becomes $0.06\%$, which is much closer to the value observed for the $\Mgal$ scenario. This lower FAP is a consequence of higher values for the lensed events combined with a slight decrease of the values for the unlensed background.

This experiment shows that there is no major consequence in making errors on the bounds of the model. However, we see that taking more stringent bounds lead to a loss in events, with some event being entirely discarded. Still, if one does not account for the lensed events with a negative $\ln(\cluModD)$, the FAP for the remaining event is decreased. On the other hand, using more conservative bounds helps retrieve all the events but leads to an increase in the FAP as the effect of the lensing statistic is weakened, making it less impactful.

\section{Conclusions}
\label{sec:Conclusions}
In this work, we have investigated how to better make the distinction between lensed and unlensed events by the usage of a rapid joint-parameter estimation pipeline and the inclusion of lensing statistics in the decision process by analyzing an extensive unlensed background with a lensed foreground. Our event pool was made to resemble as much as possible a realistic observation scenario, including changes in the PSD used to generate the noise as well as a variation in the number of detectors observing each event. This leads to an increase in the error made on the parameters, causing unlensed events to be mis-identified as lensed pairs by chance.

First, we have compared the performances of \golum~ with the results of the posterior overlap method~\citep{Haris:2018vmn}, showing that comparing the strains and ascertaining the match between all the parameters decreases significantly the false-alarm probability. Based on this, we suggest a new approach to perform online searches for strong lensing; neglecting the effect of HOMs, one could use the posterior samples from the first image obtained with traditional methods to analyze the second image under the lensed hypothesis and compare the evidence for this image with the evidence obtained for the unlensed run. This would lead to better discrimination between the lensed and unlensed events at low-latency. However, if BBH events with a large HOM content are found, this method would not be entirely trustworthy, as HOMs can impact the observed parameters and lead to bias if type II images are present\footnote{Posterior overlap suffers from the same caveat as it is performed on posteriors obtained during unlensed parameter estimation runs.}. 

Using our joint parameter estimation tool, we showed how to incorporate information on the relative magnification and time delay obtained from a lensed model without the need to re-do the parameter estimation, saving precious computational time. This can be done by reweighing the evidence obtained from the runs using the probability densities obtained from different lens catalogs. In this work, we used the results of~\citet{Haris:2018vmn}, \citet{Wierda:2021upe} and \citet{More:2021kpb} to simulate three different models for galaxy lenses. We also added four toy models representing different possible observation scenarios, such as the analysis using a galaxy-cluster lens or a change in the bounds used for the model. A census of all the models used in this work is presented in Table~\ref{tab:RecapModels}. 

\begin{figure}
    \centering
    \includegraphics[keepaspectratio, width=0.49\textwidth]{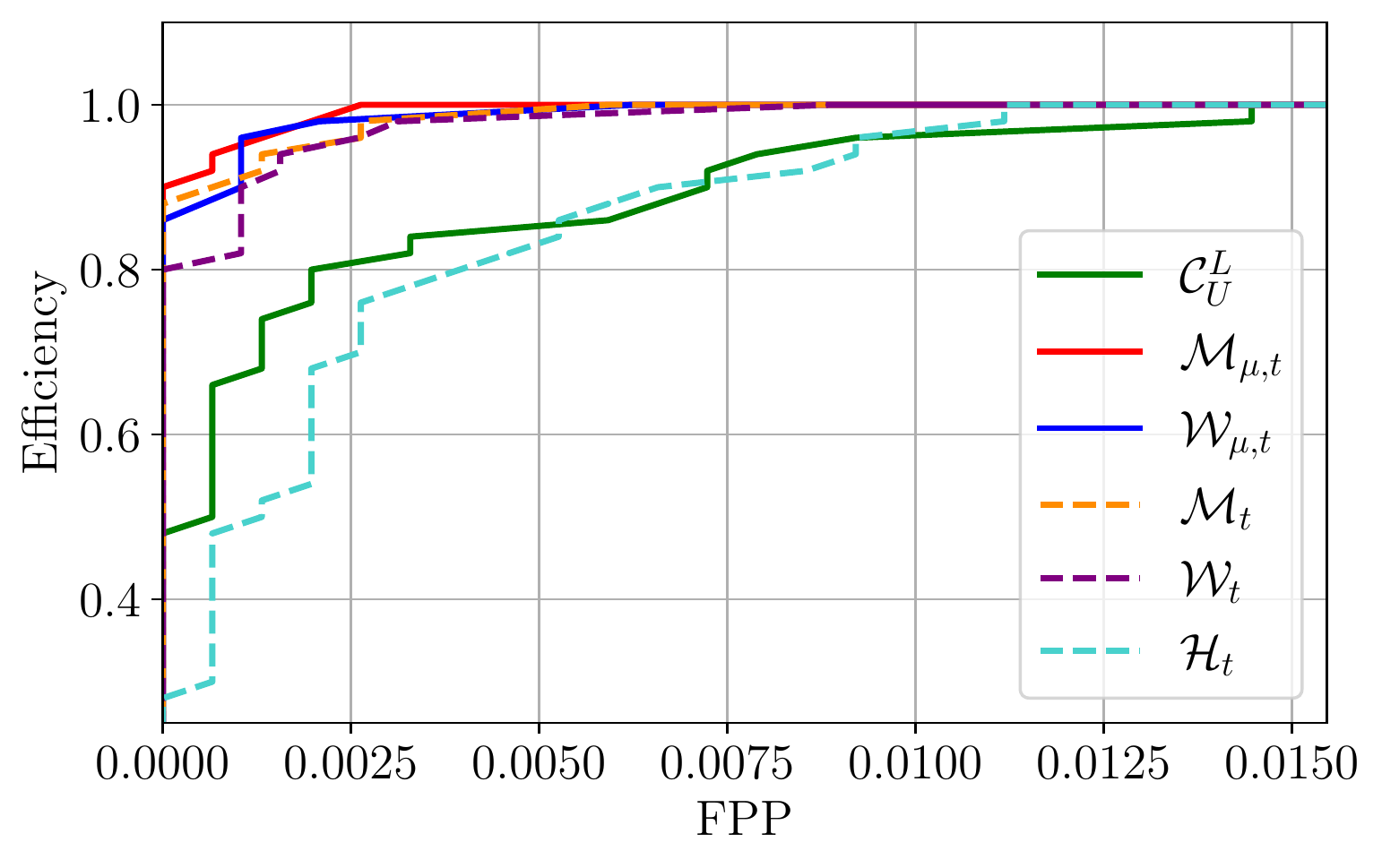}
    \caption{ROC for the exisiting catalogs with and without time delay. One sees that the inclusion of the correct model is the best possible scenario. However, using only the time delays does not lead to a major change. Another catalog using the same density profile but using shear does not drastically change the performances either. Using an erroneous density profile degrades the performances significantly, making them worse than using no model at all.}
    \label{fig:ROC_allModels}
\end{figure}

We give the FAP values obtained for all the models in Table~\ref{tab:FAPallModels} and represent the performances for the lensing catalog-based models in Fig.~\ref{fig:ROC_allModels}.

\begin{table}[t!]
    \centering
    \begin{tabular}{c|c}
    \hline
    \hline
        Model & FAP \\
    \hline
    \hline
    $\clu$ & $0.85\%$ \\
    $\Mgal$ & $0.07\%$ \\
    $\mathcal{M}_{t_{ji}}$ & $0.19\%$ \\
    $\Ugal$ & $0.1\%$ \\
    $\mathcal{W}_{t}$ & $0.21\%$ \\
    $\Rgal$ & $0.92\%$  \\
    Model A & $2.3\%$ \\
    Model B & $2.4\%$ \\
    Model C & $0.56\%$ \\
    Model D with all lensed events & $0.83\%$ \\
    Model D without discarded lensed events & $0.065\%$ \\
    \hline   
    \end{tabular}
    \caption{Summary of the FAP for all the models used in this work. There are two values given for Model D, one where we keep all the lensed events (including those having $\ln{(\cluModD)} < 0$), and the other where we do not consider the events that would not be seen as lensed (those that have $\ln{(\cluModD)} < 0$).}
    \label{tab:FAPallModels}
\end{table}

When comparing the discriminatory power between the lensed and the unlensed background for the different models, we have shown that, as expected, the best-case scenario is when one uses the correct model with both the relative magnification and time delay included. We have then also shown that having a slight change in the model (represented by the addition of shear in the model) leads to a slight increase in the FAP but does not lead to drastic modifications in the identification capabilities. This implies that minor differences between the true underlying lens model and the model chosen in our analysis will not compromise the detection efficiency significantly.

We also looked at the consequences of using only the time delay in these realistic models. This leads to a slight decrease in the efficiency compared to the case where the relative magnification is used. Indeed, some of the unlensed events have their time delay fairly compatible with the lensing distribution but not their relative magnification. Therefore, when the latter is not included, they are less well removed from the background as there is no decrease in the statistics due to the incompatibility in one of the two parameters. Nevertheless, the increase in FAP is not huge and the identification of lensing is still much easier than without using any model. We note here that the Morse factor (or phase difference) between the events has not been used in this work. However, it could also lead to more constraints and the possibility to get even better efficiency in lensing identification. The difficulty with this parameter is that the expected value is different depending on which pairs of images are seen in the lensed multiplet~\citep{More:2021kpb}. Hence, one would need to account for the uncertainty on the ordering of the observed images.

Next, we study the case where the time delay obtained from the wrong density profile, an SIS model, is assumed instead of the SIE model. Here, the identification of lensing becomes nearly impossible and the efficiency for detection is worse than the case where no model is included at all. 

This can be observed even more when looking at the results for the alternative models A and B, where we make toy models with time delays biased towards higher values. In this case, we see that some of the lensed pairs are seen as unlensed and that the confusion background is increased compared to the scenario without a lens model. This mimics the case where galaxy-scale lenses are analyzed with a cluster-scale lens model. In such a case, the identification of the lensed events would be next to impossible. This motivates the need to perform multiple lens searches in parallel, for each type of lens as the respective lens models are fairly distinct and assuming any one model can lead to missing of lenses belonging to the other type\footnote{We also expect this to be true if one analyzes galaxy cluster lensed events with a galaxy lens model.}.

Finally, with models C and D, we vary the bounds on the model to understand its impact on the detection. We found that the effect was a slight increase of the FAP when the bounds are widened while retaining all the events. On the other hand, when the bounds are tightened, we lose some of the lensed events. Lensed events with lensing parameters closer to the edge of the distribution are discarded because they get a very low probability in the lensed hypothesis. If we compute the FAP when keeping these events in the lensed foreground, it increases significantly because of the very low values of their statistics. However, by removing them from the lensed pool (which is what would be done in reality), we get a further decrease in the FAP. This indicates that the choice made for the bounds on the model will correspond to a trade-off between the robust identification or significance of the detection versus efficiency of the detection.

In conclusion, although we know that strong lensing of GW could be detected in the coming years, identifying strongly-lensed GW robustly is a real challenge. A large number of unlensed events leads to a  significant background that can lead to many false alarms. Still, there is hope. The inclusion of lensing statistics in the detection process makes the chances of correctly identifying lensing much higher. However, using a lens model does not guarantee detection of all lensed events since the efficiency of the detection is sensitive to the choice of the lens model. Therefore, our suggested approach, based on this work, is to analyze first the events without a lens model using a fast joint-parameter estimation tool, and then do a follow-up analysis for the high $\clu$ pairs using different plausible lens models for different types of lenses, not only limited to the most likely types of lenses. This would also require the development of new lens catalogs to have statistics for other lens types than galaxy lenses (More et al. 2022, in prep.). Setting up such a framework and using extended backgrounds to find the significance of the observed events should help us in the more confident identification of lensing.

\section*{Acknowledgements}
\label{sec:Acknow}
The authors are thankful to K. Haris and O.A. Hannuksela for useful discussion on the topics. 
J.J, and C.V.D.B. are supported by the research program of the Netherlands Organisation for Scientific Research (NWO). 
The authors are grateful for computational resources provided by the LIGO Laboratory and supported by the National Science Foundation Grants No. PHY-0757058 and No. PHY-0823459.

\bibliography{bibli}

\begin{thebibliography}{}
\expandafter\ifx\csname natexlab\endcsname\relax\def\natexlab#1{#1}\fi
\providecommand{\url}[1]{\href{#1}{#1}}
\providecommand{\dodoi}[1]{doi:~\href{http://doi.org/#1}{\nolinkurl{#1}}}
\providecommand{\doeprint}[1]{\href{http://ascl.net/#1}{\nolinkurl{http://ascl.net/#1}}}
\providecommand{\doarXiv}[1]{\href{https://arxiv.org/abs/#1}{\nolinkurl{https://arxiv.org/abs/#1}}}

\bibitem[{Aasi {et~al.}(2015)}]{TheLIGOScientific:2014jea}
Aasi, J., {et~al.} 2015, Class. Quant. Grav., 32, 074001,
  \dodoi{10.1088/0264-9381/32/7/074001}

\bibitem[{Abbott {et~al.}(2021{\natexlab{a}})}]{LIGOScientific:2021djp}
Abbott, R., {et~al.} 2021{\natexlab{a}}.
\newblock \doarXiv{2111.03606}

\bibitem[{Abbott {et~al.}(2021{\natexlab{b}})}]{LIGOScientific:2021sio}
---. 2021{\natexlab{b}}.
\newblock \doarXiv{2112.06861}

\bibitem[{Abbott {et~al.}(2021{\natexlab{c}})}]{LIGOScientific:2021psn}
---. 2021{\natexlab{c}}.
\newblock \doarXiv{2111.03634}

\bibitem[{Abbott {et~al.}(2021{\natexlab{d}})}]{LIGOScientific:2021izm}
---. 2021{\natexlab{d}}, Astrophys. J., 923, 14,
  \dodoi{10.3847/1538-4357/ac23db}

\bibitem[{Abbott {et~al.}(2021{\natexlab{e}})}]{LIGOScientific:2021usb}
---. 2021{\natexlab{e}}.
\newblock \doarXiv{2108.01045}

\bibitem[{Acernese {et~al.}(2015)}]{TheVirgo:2014hva}
Acernese, F., {et~al.} 2015, Class. Quant. Grav., 32, 024001,
  \dodoi{10.1088/0264-9381/32/2/024001}

\bibitem[{Akutsu {et~al.}(2019)}]{Akutsu:2018axf}
Akutsu, T., {et~al.} 2019, Nature Astron., 3, 35,
  \dodoi{10.1038/s41550-018-0658-y}

\bibitem[{{Akutsu} {et~al.}(2020){Akutsu}, {Ando}, {Arai}, {Arai}, {Araki},
  {Araya}, {Aritomi}, {Aso}, {Bae}, {Bae}, {Baiotti}, {Bajpai}, {Barton},
  {Cannon}, {Capocasa}, {Chan}, {Chen}, {Chen}, {Chen}, {Chu}, {Chu}, {Eguchi},
  {Enomoto}, {Flaminio}, {Fujii}, {Fukunaga}, {Fukushima}, {Ge}, {Hagiwara},
  {Haino}, {Hasegawa}, {Hayakawa}, {Hayama}, {Himemoto}, {Hiranuma}, {Hirata},
  {Hirose}, {Hong}, {Hsieh}, {Huang}, {Huang}, {Huang}, {Ikenoue}, {Imam},
  {Inayoshi}, {Inoue}, {Ioka}, {Itoh}, {Izumi}, {Jung}, {Jung}, {Kajita},
  {Kamiizumi}, {Kanda}, {Kang}, {Kawaguchi}, {Kawai}, {Kawasaki}, {Kim}, {Kim},
  {Kim}, {Kim}, {Kimura}, {Kita}, {Kitazawa}, {Kojima}, {Kokeyama}, {Komori},
  {Kong}, {Kotake}, {Kozakai}, {Kozu}, {Kumar}, {Kume}, {Kuo}, {Kuo},
  {Kuroyanagi}, {Kusayanagi}, {Kwak}, {Lee}, {Lee}, {Lee}, {Leonardi}, {Lin},
  {Lin}, {Lin}, {Liu}, {Luo}, {Marchio}, {Michimura}, {Mio}, {Miyakawa},
  {Miyamoto}, {Miyazaki}, {Miyo}, {Miyoki}, {Morisaki}, {Moriwaki}, {Nagano},
  {Nagano}, {Nakamura}, {Nakano}, {Nakano}, {Nakashima}, {Narikawa}, {Negishi},
  {Ni}, {Nishizawa}, {Obuchi}, {Ogaki}, {Oh}, {Oh}, {Ohashi}, {Ohishi},
  {Ohkawa}, {Okutomi}, {Oohara}, {Ooi}, {Oshino}, {Pan}, {Pang}, {Park},
  {Pe{\~n}a Arellano}, {Pinto}, {Sago}, {Saito}, {Saito}, {Sakai}, {Sakai},
  {Sakuno}, {Sato}, {Sato}, {Sawada}, {Sekiguchi}, {Sekiguchi}, {Shibagaki},
  {Shimizu}, {Shimoda}, {Shimode}, {Shinkai}, {Shishido}, {Shoda}, {Somiya},
  {Son}, {Sotani}, {Sugimoto}, {Suzuki}, {Suzuki}, {Tagoshi}, {Takahashi},
  {Takahashi}, {Takamori}, {Takano}, {Takeda}, {Takeda}, {Tanaka}, {Tanaka},
  {Tanaka}, {Tanaka}, {Tanaka}, {Tanioka}, {Tapia San Martin}, {Telada},
  {Tomaru}, {Tomigami}, {Tomura}, {Travasso}, {Trozzo}, {Tsang}, {Tsubono},
  {Tsuchida}, {Tsuzuki}, {Tuyenbayev}, {Uchikata}, {Uchiyama}, {Ueda},
  {Uehara}, {Ueno}, {Ueshima}, {Uraguchi}, {Ushiba}, {van Putten}, {Vocca},
  {Wang}, {Wu}, {Wu}, {Wu}, {Xu}, {Yamada}, {Yamamoto}, {Yamamoto}, {Yamamoto},
  {Yokogawa}, {Yokoyama}, {Yokozawa}, {Yoshioka}, {Yuzurihara}, {Zeidler},
  {Zhao}, \& {Zhu}}]{Akutsu:2020his}
{Akutsu}, T., {Ando}, M., {Arai}, K., {et~al.} 2020, arXiv e-prints,
  arXiv:2005.05574.
\newblock \doarXiv{2005.05574}

\bibitem[{Ashton {et~al.}(2019)}]{Ashton:2018jfp}
Ashton, G., {et~al.} 2019, Astrophys. J. Suppl., 241, 27,
  \dodoi{10.3847/1538-4365/ab06fc}

\bibitem[{Aso {et~al.}(2013)Aso, Michimura, Somiya, Ando, Miyakawa, Sekiguchi,
  Tatsumi, \& Yamamoto}]{Aso:2013eba}
Aso, Y., Michimura, Y., Somiya, K., {et~al.} 2013, Phys. Rev. D, 88, 043007,
  \dodoi{10.1103/PhysRevD.88.043007}

\bibitem[{Baker \& Trodden(2017)}]{Baker:2016reh}
Baker, T., \& Trodden, M. 2017, Phys. Rev. D, 95, 063512,
  \dodoi{10.1103/PhysRevD.95.063512}

\bibitem[{Barkana(1998)}]{Barkana:1998qu}
Barkana, R. 1998, Astrophys. J., 502, 531, \dodoi{10.1086/305950}

\bibitem[{Cao {et~al.}(2019)Cao, Qi, Cao, Biesiada, Li, Pan, \&
  Zhu}]{Cao:2019kgn}
Cao, S., Qi, J., Cao, Z., {et~al.} 2019, Sci. Rep., 9, 11608,
  \dodoi{10.1038/s41598-019-47616-4}

\bibitem[{Cao {et~al.}(2014)Cao, Li, \& Wang}]{Cao:2014oaa}
Cao, Z., Li, L.-F., \& Wang, Y. 2014, Phys. Rev. D, 90, 062003,
  \dodoi{10.1103/PhysRevD.90.062003}

\bibitem[{\c{C}al\i{}\c{s}kan {et~al.}(2022)\c{C}al\i{}\c{s}kan, Ezquiaga,
  Hannuksela, \& Holz}]{Caliskan:2022wbh}
\c{C}al\i{}\c{s}kan, M., Ezquiaga, J.~M., Hannuksela, O.~A., \& Holz, D.~E.
  2022.
\newblock \doarXiv{2201.04619}

\bibitem[{{Cheung} {et~al.}(2021){Cheung}, {Gais}, {Hannuksela}, \&
  {Li}}]{Cheung:2020okf}
{Cheung}, M. H.~Y., {Gais}, J., {Hannuksela}, O.~A., \& {Li}, T. G.~F. 2021,
  \mnras, 503, 3326, \dodoi{10.1093/mnras/stab579}

\bibitem[{Christian {et~al.}(2018)Christian, Vitale, \&
  Loeb}]{Christian:2018vsi}
Christian, P., Vitale, S., \& Loeb, A. 2018, Phys. Rev. D, 98, 103022,
  \dodoi{10.1103/PhysRevD.98.103022}

\bibitem[{{Dai} \& {Venumadhav}(2017)}]{Dai:2017huk}
{Dai}, L., \& {Venumadhav}, T. 2017, arXiv e-prints, arXiv:1702.04724.
\newblock \doarXiv{1702.04724}

\bibitem[{Dai {et~al.}(2017)Dai, Venumadhav, \& Sigurdson}]{Dai:2016igl}
Dai, L., Venumadhav, T., \& Sigurdson, K. 2017, \prd, 95, 044011,
  \dodoi{10.1103/PhysRevD.95.044011}

\bibitem[{Deguchi \& Watson(1986)}]{Degushi1986}
Deguchi, S., \& Watson, W.~D. 1986, Phys. Rev. D, 34, 1708,
  \dodoi{10.1103/PhysRevD.34.1708}

\bibitem[{{Ezquiaga} {et~al.}(2021){Ezquiaga}, {Holz}, {Hu}, {Lagos}, \&
  {Wald}}]{Ezquiaga:2020gdt}
{Ezquiaga}, J.~M., {Holz}, D.~E., {Hu}, W., {Lagos}, M., \& {Wald}, R.~M. 2021,
  \prd, 103, 064047, \dodoi{10.1103/PhysRevD.103.064047}

\bibitem[{Fan {et~al.}(2017)Fan, Liao, Biesiada, Piorkowska-Kurpas, \&
  Zhu}]{Fan:2016swi}
Fan, X.-L., Liao, K., Biesiada, M., Piorkowska-Kurpas, A., \& Zhu, Z.-H. 2017,
  Phys. Rev. Lett., 118, 091102, \dodoi{10.1103/PhysRevLett.118.091102}

\bibitem[{Goyal {et~al.}(2021{\natexlab{a}})Goyal, D., Kapadia, \&
  Ajith}]{Goyal:2021hxv}
Goyal, S., D., H., Kapadia, S.~J., \& Ajith, P. 2021{\natexlab{a}}, Phys. Rev.
  D, 104, 124057, \dodoi{10.1103/PhysRevD.104.124057}

\bibitem[{Goyal {et~al.}(2021{\natexlab{b}})Goyal, Haris, Mehta, \&
  Ajith}]{Goyal:2020bkm}
Goyal, S., Haris, K., Mehta, A.~K., \& Ajith, P. 2021{\natexlab{b}}, Phys. Rev.
  D, 103, 024038, \dodoi{10.1103/PhysRevD.103.024038}

\bibitem[{Hannuksela {et~al.}(2020)Hannuksela, Collett, \c{C}al\i{}\c{s}kan, \&
  Li}]{Hannuksela:2020xor}
Hannuksela, O.~A., Collett, T.~E., \c{C}al\i{}\c{s}kan, M., \& Li, T.~G. 2020,
  Mon. Not. Roy. Astron. Soc., 498, 3395, \dodoi{10.1093/mnras/staa2577}

\bibitem[{Hannuksela {et~al.}(2019)Hannuksela, Haris, Ng, Kumar, Mehta, Keitel,
  Li, \& Ajith}]{Hannuksela:2019kle}
Hannuksela, O.~A., Haris, K., Ng, K. K.~Y., {et~al.} 2019, Astrophys. J. Lett.,
  874, L2, \dodoi{10.3847/2041-8213/ab0c0f}

\bibitem[{{Haris} {et~al.}(2018){Haris}, {Mehta}, {Kumar}, {Venumadhav}, \&
  {Ajith}}]{Haris:2018vmn}
{Haris}, K., {Mehta}, A.~K., {Kumar}, S., {Venumadhav}, T., \& {Ajith}, P.
  2018, arXiv e-prints, arXiv:1807.07062.
\newblock \doarXiv{1807.07062}

\bibitem[{Hsueh {et~al.}(2018)Hsueh, Despali, Vegetti, Xu, Fassnacht, \&
  Metcalf}]{Hsueh:2017nlk}
Hsueh, J.-W., Despali, G., Vegetti, S., {et~al.} 2018, Mon. Not. Roy. Astron.
  Soc., 475, 2438, \dodoi{10.1093/mnras/stx3320}

\bibitem[{Iyer {et~al.}(2011)Iyer, Souradeep, Unnikrishnan, Dhurandhar, Raja,
  \& Sengupta}]{LigoIndia}
Iyer, B., Souradeep, T., Unnikrishnan, C., {et~al.} 2011, {LIGO-India Tech.
  rep. }, \url{https://dcc.ligo.org/LIGO-M1100296/public}

\bibitem[{Janquart {et~al.}(2021{\natexlab{a}})Janquart, Hannuksela, K., \& Van
  Den~Broeck}]{Janquart:2021qov}
Janquart, J., Hannuksela, O.~A., K., H., \& Van Den~Broeck, C.
  2021{\natexlab{a}}, Mon. Not. Roy. Astron. Soc., 506, 5430,
  \dodoi{10.1093/mnras/stab1991}

\bibitem[{Janquart {et~al.}(2021{\natexlab{b}})Janquart, Seo, Hannuksela, Li,
  \& Van Den~Broeck}]{Janquart:2021nus}
Janquart, J., Seo, E., Hannuksela, O.~A., Li, T. G.~F., \& Van Den~Broeck, C.
  2021{\natexlab{b}}, Astrophys. J. Lett., 923, L1,
  \dodoi{10.3847/2041-8213/ac3bcf}

\bibitem[{Jeffreys(2003)}]{jeffreys_2003}
Jeffreys, H. 2003, The theory of probability (Clarendon Press)

\bibitem[{{Jit Singh} {et~al.}(2018){Jit Singh}, {Li}, {Hannuksela}, {Li}, \&
  {Kim}}]{Singh:2018csp}
{Jit Singh}, A., {Li}, I. S.~C., {Hannuksela}, O.~A., {Li}, T. G.~F., \& {Kim},
  K. 2018, arXiv e-prints, arXiv:1810.07888.
\newblock \doarXiv{1810.07888}

\bibitem[{{Kim} {et~al.}(2020){Kim}, {Lee}, {Yuen}, {Akseli Hannuksela}, \&
  {Li}}]{Kim:2020xkm}
{Kim}, K., {Lee}, J., {Yuen}, R. S.~H., {Akseli Hannuksela}, O., \& {Li}, T.
  G.~F. 2020, arXiv e-prints, arXiv:2010.12093.
\newblock \doarXiv{2010.12093}

\bibitem[{{Koopmans} {et~al.}(2009){Koopmans}, {Bolton}, {Treu}, {Czoske},
  {Auger}, {Barnab{\`e}}, {Vegetti}, {Gavazzi}, {Moustakas}, \&
  {Burles}}]{Koopmans2009}
{Koopmans}, L.~V.~E., {Bolton}, A., {Treu}, T., {et~al.} 2009, \apjl, 703, L51,
  \dodoi{10.1088/0004-637X/703/1/L51}

\bibitem[{Lai {et~al.}(2018)Lai, Hannuksela, Herrera-Mart\'\i{}n, Diego,
  Broadhurst, \& Li}]{Lai:2018rto}
Lai, K.-H., Hannuksela, O.~A., Herrera-Mart\'\i{}n, A., {et~al.} 2018, Phys.
  Rev. D, 98, 083005, \dodoi{10.1103/PhysRevD.98.083005}

\bibitem[{{Li} {et~al.}(2019){Li}, {Lo}, {Sachdev}, {Chan}, {Lin}, {Li}, \&
  {Weinstein}}]{Li:2019osa}
{Li}, A. K.~Y., {Lo}, R. K.~L., {Sachdev}, S., {et~al.} 2019, arXiv e-prints,
  arXiv:1904.06020.
\newblock \doarXiv{1904.06020}

\bibitem[{Li {et~al.}(2018)Li, Mao, Zhao, \& Lu}]{Li:2018prc}
Li, S.-S., Mao, S., Zhao, Y., \& Lu, Y. 2018, Mon. Not. Roy. Astron. Soc., 476,
  2220, \dodoi{10.1093/mnras/sty411}

\bibitem[{Li {et~al.}(2019)Li, Fan, \& Gou}]{Li:2019rns}
Li, Y., Fan, X., \& Gou, L. 2019, Astrophys. J., 873, 37,
  \dodoi{10.3847/1538-4357/ab037e}

\bibitem[{Liao {et~al.}(2017)Liao, Fan, Ding, Biesiada, \& Zhu}]{Liao:2017ioi}
Liao, K., Fan, X.-L., Ding, X.-H., Biesiada, M., \& Zhu, Z.-H. 2017, Nature
  Commun., 8, 1148, \dodoi{10.1038/s41467-017-01152-9}

\bibitem[{LIGO \& Virgo(2021{\natexlab{a}})}]{GWTC2_dataRelease}
LIGO, \& Virgo. 2021{\natexlab{a}}, {GWTC-2.1: Deep Extended Catalog of Compact
  Binary Coalescences Observed by LIGO and Virgo During the First Half of the
  Third Observing Run - Parameter Estimation Data Release}

\bibitem[{LIGO \& Virgo(2021{\natexlab{b}})}]{GWTC3_dataRelease}
---. 2021{\natexlab{b}}, {GWTC-3: Compact Binary Coalescences Observed by LIGO
  and Virgo During the Second Part of the Third Observing Run — Parameter
  estimation data release}

\bibitem[{{Liu} {et~al.}(2021){Liu}, {Maga{\~n}a Hernandez}, \&
  {Creighton}}]{Liu:2020par}
{Liu}, X., {Maga{\~n}a Hernandez}, I., \& {Creighton}, J. 2021, \apj, 908, 97,
  \dodoi{10.3847/1538-4357/abd7eb}

\bibitem[{{Lo} \& {Magana Hernandez}(2021)}]{Lo:2021nae}
{Lo}, R. K.~L., \& {Magana Hernandez}, I. 2021, arXiv e-prints,
  arXiv:2104.09339.
\newblock \doarXiv{2104.09339}

\bibitem[{{Macci{\`o}} \& {Miranda}(2006)}]{Macci2006}
{Macci{\`o}}, A.~V., \& {Miranda}, M. 2006, \mnras, 368, 599,
  \dodoi{10.1111/j.1365-2966.2006.10154.x}

\bibitem[{{Mao} {et~al.}(2004){Mao}, {Jing}, {Ostriker}, \& {Weller}}]{Mao2004}
{Mao}, S., {Jing}, Y., {Ostriker}, J.~P., \& {Weller}, J. 2004, \apjl, 604, L5,
  \dodoi{10.1086/383413}

\bibitem[{McIsaac {et~al.}(2020)McIsaac, Keitel, Collett, Harry, Mozzon, Edy,
  \& Bacon}]{McIsaac:2019use}
McIsaac, C., Keitel, D., Collett, T., {et~al.} 2020, Phys. Rev. D, 102, 084031,
  \dodoi{10.1103/PhysRevD.102.084031}

\bibitem[{Meena \& Bagla(2020)}]{Meena:2019ate}
Meena, A.~K., \& Bagla, J.~S. 2020, Mon. Not. Roy. Astron. Soc., 492, 1127,
  \dodoi{10.1093/mnras/stz3509}

\bibitem[{More \& More(2021)}]{More:2021kpb}
More, A., \& More, S. 2021.
\newblock \doarXiv{2111.03091}

\bibitem[{Mukherjee {et~al.}(2021)Mukherjee, Broadhurst, Diego, Silk, \&
  Smoot}]{Mukherjee:2021qam}
Mukherjee, S., Broadhurst, T., Diego, J.~M., Silk, J., \& Smoot, G.~F. 2021,
  Mon. Not. Roy. Astron. Soc., 506, 3751, \dodoi{10.1093/mnras/stab1980}

\bibitem[{Nakamura(1998)}]{Nakamura1998}
Nakamura, T.~T. 1998, Phys. Rev. Lett., 80, 1138,
  \dodoi{10.1103/PhysRevLett.80.1138}

\bibitem[{Ng {et~al.}(2018)Ng, Wong, Broadhurst, \& Li}]{Ng:2017yiu}
Ng, K. K.~Y., Wong, K. W.~K., Broadhurst, T., \& Li, T. G.~F. 2018, Physical
  Review D, 97, 023012, \dodoi{10.1103/PhysRevD.97.023012}

\bibitem[{Oguri(2018)}]{Oguri:2018muv}
Oguri, M. 2018, Mon. Not. Roy. Astron. Soc., 480, 3842,
  \dodoi{10.1093/mnras/sty2145}

\bibitem[{{Ohanian}(1974)}]{Ohanian1974}
{Ohanian}, H.~C. 1974, International Journal of Theoretical Physics, 9, 425,
  \dodoi{10.1007/BF01810927}

\bibitem[{Pagano {et~al.}(2020)Pagano, Hannuksela, \& Li}]{Pagano:2020rwj}
Pagano, G., Hannuksela, O.~A., \& Li, T. G.~F. 2020, Astron. Astrophys., 643,
  A167, \dodoi{10.1051/0004-6361/202038730}

\bibitem[{{Pang} {et~al.}(2020){Pang}, {Hannuksela}, {Dietrich}, {Pagano}, \&
  {Harry}}]{Pang:2020qow}
{Pang}, P. T.~H., {Hannuksela}, O.~A., {Dietrich}, T., {Pagano}, G., \&
  {Harry}, I.~W. 2020, \mnras, 495, 3740, \dodoi{10.1093/mnras/staa1430}

\bibitem[{{Robertson} {et~al.}(2020){Robertson}, {Smith}, {Massey}, {Eke},
  {Jauzac}, {Bianconi}, \& {Ryczanowski}}]{Robertson:2020mfh}
{Robertson}, A., {Smith}, G.~P., {Massey}, R., {et~al.} 2020, \mnras, 495,
  3727, \dodoi{10.1093/mnras/staa1429}

\bibitem[{Ryczanowski {et~al.}(2020)Ryczanowski, Smith, Bianconi, Massey,
  Robertson, \& Jauzac}]{Ryczanowski:2020mlt}
Ryczanowski, D., Smith, G.~P., Bianconi, M., {et~al.} 2020, Mon. Not. Roy.
  Astron. Soc., 495, 1666, \dodoi{10.1093/mnras/staa1274}

\bibitem[{Schneider {et~al.}(1992)Schneider, Ehlers, \& Falco}]{Schneider:1992}
Schneider, P., Ehlers, J., \& Falco, E. 1992, Springer-Verlag

\bibitem[{Seo {et~al.}(2021)Seo, Hannuksela, \& Li}]{Seo:2021ucd}
Seo, E., Hannuksela, O.~A., \& Li, T. G.~F. 2021, in {17th International
  Conference on Topics in Astroparticle and Underground Physics}.
\newblock \doarXiv{2110.03308}

\bibitem[{Sereno {et~al.}(2011)Sereno, Jetzer, Sesana, \&
  Volonteri}]{Sereno:2011ty}
Sereno, M., Jetzer, P., Sesana, A., \& Volonteri, M. 2011, Mon. Not. Roy.
  Astron. Soc., 415, 2773, \dodoi{10.1111/j.1365-2966.2011.18895.x}

\bibitem[{Smith {et~al.}(2017)}]{Smith:2018gle}
Smith, G., {et~al.} 2017, IAU Symp., 338, 98, \dodoi{10.1017/S1743921318003757}

\bibitem[{Smith {et~al.}(2018)Smith, Jauzac, Veitch, Farr, Massey, \&
  Richard}]{Smith:2017mqu}
Smith, G.~P., Jauzac, M., Veitch, J., {et~al.} 2018, Mon. Not. Roy. Astron.
  Soc., 475, 3823, \dodoi{10.1093/mnras/sty031}

\bibitem[{{Smith} {et~al.}(2019){Smith}, {Robertson}, {Bianconi}, \&
  {Jauzac}}]{Smith:2019dis}
{Smith}, G.~P., {Robertson}, A., {Bianconi}, M., \& {Jauzac}, M. 2019, arXiv
  e-prints, arXiv:1902.05140.
\newblock \doarXiv{1902.05140}

\bibitem[{Somiya(2012)}]{Somiya:2011np}
Somiya, K. 2012, Class. Quant. Grav., 29, 124007,
  \dodoi{10.1088/0264-9381/29/12/124007}

\bibitem[{{Speagle}(2020)}]{Speagle:2020}
{Speagle}, J.~S. 2020, \mnras, 493, 3132, \dodoi{10.1093/mnras/staa278}

\bibitem[{Takahashi \& Nakamura(2003)}]{Takahashi:2003ix}
Takahashi, R., \& Nakamura, T. 2003, Astrophys. J., 595, 1039,
  \dodoi{10.1086/377430}

\bibitem[{{The LIGO Scientific Collaboration} {et~al.}(2021){The LIGO
  Scientific Collaboration}, {the Virgo Collaboration}, {the KAGRA
  Collaboration}, {Abbott}, {Abe}, {Acernese}, {Ackley}, {Adhikari},
  {Adhikari}, {Adkins}, {Adya}, {Affeldt}, {Agarwal}, {Agathos}, {Agatsuma},
  {Aggarwal}, {Aguiar}, {Aiello}, {Ain}, {Ajith}, {Akutsu}, {Albanesi},
  {Alfaidi}, {Allocca}, {Altin}, {Amato}, {Anand}, {Anand}, {Ananyeva},
  {Anderson}, {Anderson}, {Ando}, {Andrade}, {Andres}, {Andr{\'e}s-Carcasona},
  {Andri{\'c}}, {Angelova}, {Ansoldi}, {Antelis}, {Antier}, {Apostolatos},
  {Appavuravther}, {Appert}, {Apple}, {Arai}, {Araya}, {Araya}, {Areeda},
  {Ar{\`e}ne}, {Aritomi}, {Arnaud}, {Arogeti}, {Aronson}, {Arun}, {Asada},
  {Asali}, {Ashton}, {Aso}, {Assiduo}, {Assis de Souza Melo}, {Aston},
  {Astone}, {Aubin}, {AultONeal}, {Austin}, {Babak}, {Badaracco}, {Bader},
  {Badger}, {Bae}, {Bae}, {Baer}, {Bagnasco}, {Bai}, {Baird}, {Bajpai}, {Baka},
  {Ball}, {Ballardin}, {Ballmer}, {Balsamo}, {Baltus}, {Banagiri}, {Banerjee},
  {Bankar}, {Barayoga}, {Barbieri}, {Barbieri}, {Barish}, {Barker}, {Barneo},
  {Barone}, {Barr}, {Barsotti}, {Barsuglia}, {Barta}, {Bartlett}, {Barton},
  {Bartos}, {Basak}, {Bassiri}, {Basti}, {Bawaj}, {Bayley}, {Bazzan}, {Becher},
  {B{\'e}csy}, {Bedakihale}, {Beirnaert}, {Bejger}, {Belahcene}, {Benedetto},
  {Beniwal}, {Benjamin}, {Bennett}, {Bentley}, {BenYaala}, {Bera}, {Berbel},
  {Bergamin}, {Berger}, {Bernuzzi}, {Berry}, {Bersanetti}, {Bertolini},
  {Betzwieser}, {Beveridge}, {Bhandare}, {Bhandari}, {Bhardwaj}, {Bhatt},
  {Bhattacharjee}, {Bhaumik}, {Bianchi}, {Bilenko}, {Billingsley}, {Bini},
  {Birney}, {Birnholtz}, {Biscans}, {Bischi}, {Biscoveanu}, {Bisht}, {Biswas},
  {Bitossi}, {Bizouard}, {Blackburn}, {Blair}, {Blair}, {Blair}, {Bobba},
  {Bode}, {Bo{\"e}r}, {Bogaert}, {Boldrini}, {Bolingbroke}, {Bonavena},
  {Bondu}, {Bonilla}, {Bonnand}, {Booker}, {Boom}, {Bork}, {Boschi}, {Bose},
  {Bose}, {Bossilkov}, {Boudart}, {Bouffanais}, {Bozzi}, {Bradaschia}, {Brady},
  {Bramley}, {Branch}, {Branchesi}, {Brau}, {Breschi}, {Briant}, {Briggs},
  {Brillet}, {Brinkmann}, {Brockill}, {Brooks}, {Brooks}, {Brown}, {Brunett},
  {Bruno}, {Bruntz}, {Bryant}, {Bucci}, {Bulik}, {Bulten}, {Buonanno},
  {Burtnyk}, {Buscicchio}, {Buskulic}, {Buy}, {Byer}, {Cabourn Davies},
  {Cabras}, {Cabrita}, {Cadonati}, {Caesar}, {Cagnoli}, {Cahillane},
  {Calder{\'o}n Bustillo}, {Callaghan}, {Callister}, {Calloni}, {Cameron},
  {Camp}, {Canepa}, {Canevarolo}, {Cannavacciuolo}, {Cannon}, {Cao}, {Cao},
  {Capocasa}, {Capote}, {Carapella}, {Carbognani}, {Carlassara}, {Carlin},
  {Carney}, {Carpinelli}, {Carrillo}, {Carullo}, {Carver}, {Casanueva Diaz},
  {Casentini}, {Castaldi}, {Caudill}, {Cavagli{\`a}}, {Cavalier}, {Cavalieri},
  {Cella}, {Cerd{\'a}-Dur{\'a}n}, {Cesarini}, {Chaibi}, {Chalathadka
  Subrahmanya}, {Champion}, {Chan}, {Chan}, {Chan}, {Chan}, {Chan}, {Chandra},
  {Chang}, {Chanial}, {Chao}, {Chapman-Bird}, {Charlton}, {Chase},
  {Chassande-Mottin}, {Chatterjee}, {Chatterjee}, {Chatterjee}, {Chaturvedi},
  {Chaty}, {Chatziioannou}, {Chen}, {Chen}, {Chen}, {Chen}, {Chen}, {Chen},
  {Chen}, {Chen}, {Chen}, {Cheng}, {Cheong}, {Cheung}, {Chia}, {Chiadini},
  {Chiang}, {Chiarini}, {Chierici}, {Chincarini}, {Chiofalo}, {Chiummo},
  {Choudhary}, {Choudhary}, {Christensen}, {Chu}, {Chu}, {Chua}, {Chung},
  {Ciani}, {Ciecielag}, {Cie{\'s}lar}, {Cifaldi}, {Ciobanu}, {Ciolfi},
  {Cipriano}, {Clara}, {Clark}, {Clearwater}, {Clesse}, {Cleva}, {Coccia},
  {Codazzo}, {Cohadon}, {Cohen}, {Colleoni}, {Collette}, {Colombo}, {Colpi},
  {Compton}, {Constancio}, {Conti}, {Cooper}, {Corban}, {Corbitt},
  {Cordero-Carri{\'o}n}, {Corezzi}, {Corley}, {Cornish}, {Corre}, {Corsi},
  {Cortese}, {Costa}, {Cotesta}, {Cottingham}, {Coughlin}, {Coulon},
  {Countryman}, {Cousins}, {Couvares}, {Coward}, {Cowart}, {Coyne}, {Coyne},
  {Creighton}, {Creighton}, {Criswell}, {Croquette}, {Crowder}, {Cudell},
  {Cullen}, {Cumming}, {Cummings}, {Cunningham}, {Cuoco}, {Cury{\l}o},
  {Dabadie}, {Dal Canton}, {Dall'Osso}, {D{\'a}lya}, {Dana}, {D'Angelo},
  {Danilishin}, {D'Antonio}, {Danzmann}, {Darsow-Fromm}, {Dasgupta}, {Datrier},
  {Datta}, {Datta}, {Dattilo}, {Dave}, {Davier}, {Davis}, {Davis}, {Daw}, {de
  Alarc{\'o}n}, {Dean}, {DeBra}, {Deenadayalan}, {Degallaix}, {De Laurentis},
  {Del{\'e}glise}, {Del Favero}, {De Lillo}, {De Lillo}, {Dell'Aquila}, {Del
  Pozzo}, {DeMarchi}, {De Matteis}, {D'Emilio}, {Demos}, {Dent}, {Depasse}, {De
  Pietri}, {De Rosa}, {De Rossi}, {DeSalvo}, {De Simone}, {Dhurandhar},
  {D{\'\i}az}, {Didio}, {Dietrich}, {Di Fiore}, {Di Fronzo}, {Di Giorgio}, {Di
  Giovanni}, {Di Giovanni}, {Di Girolamo}, {Di Lieto}, {Di Michele}, {Ding},
  {Di Pace}, {Di Palma}, {Di Renzo}, {Divakarla}, {Dmitriev}, {Doctor},
  {Donahue}, {D'Onofrio}, {Donovan}, {Dooley}, {Doravari}, {Drago}, {Driggers},
  {Drori}, {Ducoin}, {Dupej}, {Dupletsa}, {Durante}, {D'Urso}, {Duverne},
  {Dwyer}, {Eassa}, {Easter}, {Ebersold}, {Eckhardt}, {Eddolls}, {Edelman},
  {Edo}, {Edy}, {Effler}, {Eguchi}, {Eichholz}, {Eikenberry}, {Eisenmann},
  {Eisenstein}, {Ejlli}, {Engelby}, {Enomoto}, {Errico}, {Essick},
  {Estell{\'e}s}, {Estevez}, {Etienne}, {Etzel}, {Evans}, {Evans},
  {Evstafyeva}, {Ewing}, {Fabrizi}, {Faedi}, {Fafone}, {Fair}, {Fairhurst},
  {Fan}, {Farah}, {Farinon}, {Farr}, {Farr}, {Fauchon-Jones}, {Favaro},
  {Favata}, {Fays}, {Fazio}, {Feicht}, {Fejer}, {Fenyvesi}, {Ferguson},
  {Fernandez-Galiana}, {Ferrante}, {Ferreira}, {Fidecaro}, {Figura}, {Fiori},
  {Fiori}, {Fishbach}, {Fisher}, {Fittipaldi}, {Fiumara}, {Flaminio}, {Floden},
  {Fong}, {Font}, {Fornal}, {Forsyth}, {Franke}, {Frasca}, {Frasconi}, {Freed},
  {Frei}, {Freise}, {Freitas}, {Frey}, {Fritschel}, {Frolov}, {Fronz{\'e}},
  {Fujii}, {Fujikawa}, {Fujimoto}, {Fulda}, {Fyffe}, {Gabbard}, {Gadre},
  {Gair}, {Gais}, {Galaudage}, {Gamba}, {Ganapathy}, {Ganguly}, {Gao},
  {Gaonkar}, {Garaventa}, {Garc{\'\i}a N{\'u}{\~n}ez},
  {Garc{\'\i}a-Quir{\'o}s}, {Garufi}, {Gateley}, {Gayathri}, {Ge}, {Gemme},
  {Gennai}, {George}, {Gerberding}, {Gergely}, {Gewecke}, {Ghonge}, {Ghosh},
  {Ghosh}, {Ghosh}, {Ghosh}, {Ghosh}, {Giacomazzo}, {Giacoppo}, {Giaime},
  {Giardina}, {Gibson}, {Gier}, {Giesler}, {Giri}, {Gissi}, {Gkaitatzis},
  {Glanzer}, {Gleckl}, {Godwin}, {Goetz}, {Goetz}, {Gohlke}, {Golomb},
  {Goncharov}, {Gonz{\'a}lez}, {Gosselin}, {Gouaty}, {Gould}, {Goyal}, {Grace},
  {Grado}, {Graham}, {Granata}, {Granata}, {Grant}, {Gras}, {Grassia}, {Gray},
  {Gray}, {Greco}, {Green}, {Green}, {Gretarsson}, {Gretarsson}, {Griffith},
  {Griffiths}, {Griggs}, {Grignani}, {Grimaldi}, {Grimes}, {Grimm}, {Grote},
  {Grunewald}, {Gruning}, {Gruson}, {Guerra}, {Guidi}, {Guimaraes},
  {Guix{\'e}}, {Gulati}, {Gunny}, {Guo}, {Guo}, {Gupta}, {Gupta}, {Gupta},
  {Gupta}, {Gupta}, {Gustafson}, {Guzman}, {Ha}, {Hadiputrawan}, {Haegel},
  {Haino}, {Halim}, {Hall}, {Hamilton}, {Hammond}, {Han}, {Haney}, {Hanks},
  {Hanna}, {Hannam}, {Hannuksela}, {Hansen}, {Hansen}, {Hanson}, {Harder},
  {Haris}, {Harms}, {Harry}, {Harry}, {Hartwig}, {Hasegawa}, {Haskell},
  {Haster}, {Hathaway}, {Hattori}, {Haughian}, {Hayakawa}, {Hayama}, {Hayes},
  {Healy}, {Heidmann}, {Heidt}, {Heintze}, {Heinze}, {Heinzel}, {Heitmann},
  {Hellman}, {Hello}, {Helmling-Cornell}, {Hemming}, {Hendry}, {Heng},
  {Hennes}, {Hennig}, {Hennig}, {Henshaw}, {Hernandez}, {Hernandez Vivanco},
  {Heurs}, {Hewitt}, {Higginbotham}, {Hild}, {Hill}, {Himemoto}, {Hines},
  {Hirata}, {Hirose}, {Ho}, {Hochheim}, {Hofman}, {Hohmann}, {Holcomb},
  {Holland}, {Hollows}, {Holmes}, {Holt}, {Holz}, {Hong}, {Hough}, {Hourihane},
  {Howell}, {Hoy}, {Hoyland}, {Hreibi}, {Hsieh}, {Hsieh}, {Hsiung}, {Hsu},
  {Huang}, {Huang}, {Huang}, {Huang}, {Huang}, {Huang}, {H{\"u}bner},
  {Huddart}, {Hughey}, {Hui}, {Hui}, {Husa}, {Huttner}, {Huxford},
  {Huynh-Dinh}, {Ide}, {Idzkowski}, {Iess}, {Inayoshi}, {Inoue}, {Iosif},
  {Isi}, {Isleif}, {Ito}, {Itoh}, {Iyer}, {JaberianHamedan}, {Jacqmin},
  {Jacquet}, {Jadhav}, {Jadhav}, {Jain}, {James}, {Jan}, {Jani}, {Janquart},
  {Janssens}, {Janthalur}, {Jaranowski}, {Jariwala}, {Jaume}, {Jenkins},
  {Jenner}, {Jeon}, {Jia}, {Jiang}, {Jin}, {Johns}, {Johnston}, {Jones},
  {Jones}, {Jones}, {Jones}, {Joshi}, {Ju}, {Jue}, {Jung}, {Jung}, {Junker},
  {Juste}, {Kaihotsu}, {Kajita}, {Kakizaki}, {Kalaghatgi}, {Kalogera}, {Kamai},
  {Kamiizumi}, {Kanda}, {Kandhasamy}, {Kang}, {Kanner}, {Kao}, {Kapadia},
  {Kapasi}, {Karathanasis}, {Karki}, {Kashyap}, {Kasprzack}, {Kastaun}, {Kato},
  {Katsanevas}, {Katsavounidis}, {Katzman}, {Kaur}, {Kawabe}, {Kawaguchi},
  {K{\'e}f{\'e}lian}, {Keitel}, {Key}, {Khadka}, {Khalili}, {Khan}, {Khanam},
  {Khazanov}, {Khetan}, {Khursheed}, {Kijbunchoo}, {Kim}, {Kim}, {Kim}, {Kim},
  {Kim}, {Kim}, {Kim}, {Kimball}, {Kimura}, {Kinley-Hanlon}, {Kirchhoff},
  {Kissel}, {Klimenko}, {Klinger}, {Knee}, {Knowles}, {Knust}, {Knyazev},
  {Kobayashi}, {Koch}, {Koekoek}, {Kohri}, {Kokeyama}, {Koley}, {Kolitsidou},
  {Kolstein}, {Komori}, {Kondrashov}, {Kong}, {Kontos}, {Koper}, {Korobko},
  {Kovalam}, {Koyama}, {Kozak}, {Kozakai}, {Kringel}, {Krishnendu},
  {Kr{\'o}lak}, {Kuehn}, {Kuei}, {Kuijer}, {Kulkarni}, {Kumar}, {Kumar},
  {Kumar}, {Kumar}, {Kume}, {Kuns}, {Kuromiya}, {Kuroyanagi}, {Kwak},
  {Lacaille}, {Lagabbe}, {Laghi}, {Lalande}, {Lalleman}, {Lam}, {Lamberts},
  {Landry}, {Lane}, {Lang}, {Lange}, {Lantz}, {La Rosa}, {Lartaux-Vollard},
  {Lasky}, {Laxen}, {Lazzarini}, {Lazzaro}, {Leaci}, {Leavey}, {LeBohec},
  {Lecoeuche}, {Lee}, {Lee}, {Lee}, {Lee}, {Lee}, {Legred}, {Lehmann},
  {Lema{\^\i}tre}, {Lenti}, {Leonardi}, {Leonova}, {Leroy}, {Letendre},
  {Levesque}, {Levin}, {Leviton}, {Leyde}, {Li}, {Li}, {Li}, {Li}, {Li}, {Li},
  {Li}, {Lin}, {Lin}, {Lin}, {Lin}, {Lin}, {Lin}, {Linde}, {Linker}, {Linley},
  {Littenberg}, {Liu}, {Liu}, {Liu}, {Liu}, {Llamas}, {Lo}, {Lo}, {London},
  {Longo}, {Lopez}, {Lopez Portilla}, {Lorenzini}, {Loriette}, {Lormand},
  {Losurdo}, {Lott}, {Lough}, {Lousto}, {Lovelace}, {Lucaccioni}, {L{\"u}ck},
  {Lumaca}, {Lundgren}, {Luo}, {Lynam}, {Ma'arif}, {Macas}, {Machtinger},
  {MacInnis}, {Macleod}, {MacMillan}, {Macquet}, {Maga{\~n}a Hernandez},
  {Magazz{\`u}}, {Magee}, {Maggiore}, {Magnozzi}, {Mahesh}, {Majorana},
  {Maksimovic}, {Maliakal}, {Malik}, {Man}, {Mandic}, {Mangano}, {Mansell},
  {Manske}, {Mantovani}, {Mapelli}, {Marchesoni}, {Mar{\'\i}n Pina}, {Marion},
  {Mark}, {M{\'a}rka}, {M{\'a}rka}, {Markakis}, {Markosyan}, {Markowitz},
  {Maros}, {Marquina}, {Marsat}, {Martelli}, {Martin}, {Martin}, {Martinez},
  {Martinez}, {Martinez}, {Martinovic}, {Martynov}, {Marx}, {Masalehdan},
  {Mason}, {Massera}, {Masserot}, {Masso-Reid}, {Mastrogiovanni}, {Matas},
  {Mateu-Lucena}, {Matichard}, {Matiushechkina}, {Mavalvala}, {McCann},
  {McCarthy}, {McClelland}, {McClincy}, {McCormick}, {McCuller}, {McGhee},
  {McGuire}, {McIsaac}, {McIver}, {McRae}, {McWilliams}, {Meacher}, {Mehmet},
  {Mehta}, {Meijer}, {Melatos}, {Melchor}, {Mendell}, {Menendez-Vazquez},
  {Menoni}, {Mercer}, {Mereni}, {Merfeld}, {Merilh}, {Merritt}, {Merzougui},
  {Meshkov}, {Messenger}, {Messick}, {Meyers}, {Meylahn}, {Mhaske}, {Miani},
  {Miao}, {Michaloliakos}, {Michel}, {Michimura}, {Middleton}, {Mihaylov},
  {Milano}, {Miller}, {Miller}, {Miller}, {Millhouse}, {Mills}, {Milotti},
  {Minenkov}, {Mio}, {Mir}, {Miravet-Ten{\'e}s}, {Mishkin}, {Mishra}, {Mishra},
  {Mistry}, {Mitra}, {Mitrofanov}, {Mitselmakher}, {Mittleman}, {Miyakawa},
  {Miyo}, {Miyoki}, {Mo}, {Modafferi}, {Moguel}, {Mogushi}, {Mohapatra},
  {Mohite}, {Molina}, {Molina-Ruiz}, {Mondin}, {Montani}, {S}, {More}, {Moore},
  {Moragues}, {Moraru}, {Morawski}, {More}, {Moreno}, {Moreno}, {Mori},
  {Morisaki}, {Morisue}, {Moriwaki}, {Mours}, {Mow-Lowry}, {Mozzon},
  {Muciaccia}, {Mukherjee}, {Mukherjee}, {Mukherjee}, {Mukherjee}, {Mukherjee},
  {Mukund}, {Mullavey}, {Munch}, {Mu{\~n}iz}, {Murray}, {Musenich}, {Muusse},
  {Nadji}, {Nagano}, {Nagar}, {Nakamura}, {Nakano}, {Nakano}, {Nakayama},
  {Napolano}, {Nardecchia}, {Narikawa}, {Narola}, {Naticchioni}, {Nayak},
  {Nayak}, {Neil}, {Neilson}, {Nelson}, {Nelson}, {Nery}, {Neubauer},
  {Neunzert}, {Ng}, {Ng}, {Nguyen}, {Nguyen}, {Nguyen}, {Nguyen Quynh}, {Ni},
  {Ni}, {Nichols}, {Nishimoto}, {Nishizawa}, {Nissanke}, {Nitoglia}, {Nocera},
  {Norman}, {North}, {Nozaki}, {Nurbek}, {Nuttall}, {Obayashi}, {Oberling},
  {O'Brien}, {O'Dell}, {Oelker}, {Ogaki}, {Oganesyan}, {Oh}, {Oh}, {Oh},
  {Ohashi}, {Ohashi}, {Ohkawa}, {Ohme}, {Ohta}, {Okada}, {Okutani}, {Olivetto},
  {Oohara}, {Oram}, {O'Reilly}, {Ormiston}, {Ormsby}, {O'Shaughnessy},
  {O'Shea}, {Oshino}, {Ossokine}, {Osthelder}, {Otabe}, {Ottaway}, {Overmier},
  {Pace}, {Pagano}, {Pagano}, {Page}, {Pagliaroli}, {Pai}, {Pai}, {Pal},
  {Palamos}, {Palashov}, {Palomba}, {Pan}, {Pan}, {Panda}, {Pang}, {Pankow},
  {Pannarale}, {Pant}, {Panther}, {Paoletti}, {Paoli}, {Paolone}, {Pappas},
  {Parisi}, {Park}, {Park}, {Parker}, {Pascucci}, {Pasqualetti}, {Passaquieti},
  {Passuello}, {Patel}, {Pathak}, {Patricelli}, {Patron}, {Paul}, {Payne},
  {Pedraza}, {Pedurand}, {Pegoraro}, {Pele}, {Pe{\~n}a Arellano}, {Penano},
  {Penn}, {Perego}, {Pereira}, {Pereira}, {Perez}, {P{\'e}rigois}, {Perkins},
  {Perreca}, {Perri{\`e}s}, {Pesios}, {Petermann}, {Petterson}, {Pfeiffer},
  {Pham}, {Pham}, {Phukon}, {Phurailatpam}, {Piccinni}, {Pichot}, {Piendibene},
  {Piergiovanni}, {Pierini}, {Pierro}, {Pillant}, {Pillas}, {Pilo}, {Pinard},
  {Pineda-Bosque}, {Pinto}, {Pinto}, {Piotrzkowski}, {Piotrzkowski}, {Pirello},
  {Pitkin}, {Placidi}, {Placidi}, {Planas}, {Plastino}, {Pluchar}, {Poggiani},
  {Polini}, {Pong}, {Ponrathnam}, {Porter}, {Poulton}, {Poverman}, {Powell},
  {Pracchia}, {Pradier}, {Prajapati}, {Prasai}, {Prasanna}, {Pratten},
  {Principe}, {Prodi}, {Prokhorov}, {Prosposito}, {Prudenzi}, {Puecher},
  {Punturo}, {Puosi}, {Puppo}, {P{\"u}rrer}, {Qi}, {Quartey}, {Quetschke},
  {Quinonez}, {Quitzow-James}, {Raab}, {Raaijmakers}, {Radkins}, {Radulesco},
  {Raffai}, {Rail}, {Raja}, {Rajan}, {Ramirez}, {Ramirez}, {Ramos-Buades},
  {Rana}, {Rapagnani}, {Ray}, {Raymond}, {Raza}, {Razzano}, {Read}, {Rees},
  {Regimbau}, {Rei}, {Reid}, {Reid}, {Reitze}, {Relton}, {Renzini}, {Rettegno},
  {Revenu}, {Reza}, {Rezac}, {Ricci}, {Richards}, {Richardson}, {Richardson},
  {Riemenschneider}, {Riles}, {Rinaldi}, {Rink}, {Robertson}, {Robie},
  {Robinet}, {Rocchi}, {Rodriguez}, {Rolland}, {Rollins}, {Romanelli},
  {Romano}, {Romel}, {Romero}, {Romero-Shaw}, {Romie}, {Ronchini}, {Rosa},
  {Rose}, {Rosi{\'n}ska}, {Ross}, {Rowan}, {Rowlinson}, {Roy}, {Roy}, {Roy},
  {Rozza}, {Ruggi}, {Ruiz-Rocha}, {Ryan}, {Sachdev}, {Sadecki}, {Sadiq},
  {Saha}, {Saito}, {Sakai}, {Sakellariadou}, {Sakon}, {Salafia},
  {Salces-Carcoba}, {Salconi}, {Saleem}, {Salemi}, {Samajdar}, {Sanchez},
  {Sanchez}, {Sanchez}, {Sanchis-Gual}, {Sanders}, {Sanuy}, {Saravanan},
  {Sarin}, {Sassolas}, {Satari}, {Sathyaprakash}, {Sauter}, {Savage}, {Savant},
  {Sawada}, {Sawant}, {Sayah}, {Schaetzl}, {Scheel}, {Scheuer}, {Schiworski},
  {Schmidt}, {Schmidt}, {Schnabel}, {Schneewind}, {Schofield}, {Sch{\"o}nbeck},
  {Schulte}, {Schutz}, {Schwartz}, {Scott}, {Scott}, {Seglar-Arroyo},
  {Sekiguchi}, {Sellers}, {Sengupta}, {Sentenac}, {Seo}, {Sequino}, {Sergeev},
  {Setyawati}, {Shaffer}, {Shahriar}, {Shaikh}, {Shams}, {Shao}, {Sharma},
  {Sharma}, {Shawhan}, {Shcheblanov}, {Sheela}, {Shikano}, {Shikauchi},
  {Shimizu}, {Shimode}, {Shinkai}, {Shishido}, {Shoda}, {Shoemaker},
  {Shoemaker}, {ShyamSundar}, {Sieniawska}, {Sigg}, {Silenzi}, {Singer},
  {Singh}, {Singh}, {Singh}, {Singha}, {Sintes}, {Sipala}, {Skliris},
  {Slagmolen}, {Slaven-Blair}, {Smetana}, {Smith}, {Smith}, {Smith},
  {Soldateschi}, {Somala}, {Somiya}, {Song}, {Soni}, {Sordini}, {Sorrentino},
  {Sorrentino}, {Soulard}, {Souradeep}, {Sowell}, {Spagnuolo}, {Spencer},
  {Spera}, {Spinicelli}, {Srivastava}, {Srivastava}, {Staats}, {Stachie},
  {Stachurski}, {Steer}, {Steinlechner}, {Steinlechner}, {Stergioulas},
  {Stops}, {Stover}, {Strain}, {Strang}, {Stratta}, {Strong}, {Strunk},
  {Sturani}, {Stuver}, {Suchenek}, {Sudhagar}, {Sudhir}, {Sugimoto}, {Suh},
  {Sullivan}, {Summerscales}, {Sun}, {Sunil}, {Sur}, {Suresh}, {Sutton},
  {Suzuki}, {Suzuki}, {Suzuki}, {Swinkels}, {Szczepa{\'n}czyk}, {Szewczyk},
  {Tacca}, {Tagoshi}, {Tait}, {Takahashi}, {Takahashi}, {Takano}, {Takeda},
  {Takeda}, {Talbot}, {Talbot}, {Tamanini}, {Tanaka}, {Tanaka}, {Tanaka},
  {Tanasijczuk}, {Tanioka}, {Tanner}, {Tao}, {Tao}, {Tapia}, {Tapia San
  Mart{\'\i}n}, {Taranto}, {Taruya}, {Tasson}, {Tenorio}, {Terhune},
  {Terkowski}, {Thirugnanasambandam}, {Thomas}, {Thomas}, {Thompson},
  {Thompson}, {Thondapu}, {Thorne}, {Thrane}, {Tiwari}, {Tiwari}, {Tiwari},
  {Toivonen}, {Tolley}, {Tomaru}, {Tomura}, {Tonelli}, {Tornasi},
  {Torres-Forn{\'e}}, {Torrie}, {Tosta e Melo}, {T{\"o}yr{\"a}}, {Trapananti},
  {Travasso}, {Traylor}, {Trevor}, {Tringali}, {Tripathee}, {Troiano},
  {Trovato}, {Trozzo}, {Trudeau}, {Tsai}, {Tsang}, {Tsang}, {Tsao}, {Tse},
  {Tso}, {Tsuchida}, {Tsukada}, {Tsuna}, {Tsutsui}, {Turbang}, {Turconi},
  {Turski}, {Tuyenbayev}, {Ubhi}, {Uchikata}, {Uchiyama}, {Udall}, {Ueda},
  {Uehara}, {Ueno}, {Ueshima}, {Unnikrishnan}, {Urban}, {Ushiba}, {Utina},
  {Vajente}, {Vajpeyi}, {Valdes}, {Valentini}, {Valsan}, {van Bakel}, {van
  Beuzekom}, {van Dael}, {van den Brand}, {Van Den Broeck}, {Vander-Hyde}, {van
  Haevermaet}, {van Heijningen}, {van Putten}, {van Remortel}, {Vardaro},
  {Vargas}, {Varma}, {Vas{\'u}th}, {Vecchio}, {Vedovato}, {Veitch}, {Veitch},
  {Venneberg}, {Venugopalan}, {Verkindt}, {Verma}, {Verma}, {Vermeulen},
  {Veske}, {Vetrano}, {Vicer{\'e}}, {Vidyant}, {Viets}, {Vijaykumar},
  {Villa-Ortega}, {Vinet}, {Virtuoso}, {Vitale}, {Vocca}, {von Reis}, {von
  Wrangel}, {Vorvick}, {Vyatchanin}, {Wade}, {Wade}, {Wagner}, {Walet},
  {Walker}, {Wallace}, {Wallace}, {Wang}, {Wang}, {Wang}, {Ward}, {Warner},
  {Was}, {Washimi}, {Washington}, {Watchi}, {Weaver}, {Weaving}, {Webster},
  {Weinert}, {Weinstein}, {Weiss}, {Weller}, {Weller}, {Wellmann}, {Wen},
  {We{\ss}els}, {Wette}, {Whelan}, {White}, {Whiting}, {Whittle}, {Wilken},
  {Williams}, {Williams}, {Williamson}, {Willis}, {Willke}, {Wilson}, {Wipf},
  {Wlodarczyk}, {Woan}, {Woehler}, {Wofford}, {Wong}, {Wong}, {Wright}, {Wu},
  {Wu}, {Wu}, {Wysocki}, {Xiao}, {Yamada}, {Yamamoto}, {Yamamoto}, {Yamamoto},
  {Yamashita}, {Yamazaki}, {Yang}, {Yang}, {Yang}, {Yang}, {Yang}, {Yang},
  {Yap}, {Yeeles}, {Yeh}, {Yelikar}, {Ying}, {Yokoyama}, {Yokozawa}, {Yoo},
  {Yoshioka}, {Yu}, {Yu}, {Yuzurihara}, {Zadro{\.z}ny}, {Zanolin}, {Zeidler},
  {Zelenova}, {Zendri}, {Zevin}, {Zhan}, {Zhang}, {Zhang}, {Zhang}, {Zhang},
  {Zhang}, {Zhang}, {Zhao}, {Zhao}, {Zhao}, {Zhao}, {Zhou}, {Zhou}, {Zhu},
  {Zhu}, {Zimmerman}, {Zucker}, \& {Zweizig}}]{GWTC3cosmology}
{The LIGO Scientific Collaboration}, {the Virgo Collaboration}, {the KAGRA
  Collaboration}, {et~al.} 2021, arXiv e-prints, arXiv:2111.03604.
\newblock \doarXiv{2111.03604}

\bibitem[{Vijaykumar {et~al.}(2022)Vijaykumar, Mehta, \&
  Ganguly}]{Vijaykumar:2022dlp}
Vijaykumar, A., Mehta, A.~K., \& Ganguly, A. 2022.
\newblock \doarXiv{2202.06334}

\bibitem[{Wang {et~al.}(2021)Wang, Lo, Li, \& Chen}]{Wang:2021kzt}
Wang, Y., Lo, R. K.~L., Li, A. K.~Y., \& Chen, Y. 2021, Phys. Rev. D, 103,
  104055, \dodoi{10.1103/PhysRevD.103.104055}

\bibitem[{Wang {et~al.}(1996)Wang, Stebbins, \& Turner}]{Wang:1996as}
Wang, Y., Stebbins, A., \& Turner, E.~L. 1996, Phys. Rev. Lett., 77, 2875,
  \dodoi{10.1103/PhysRevLett.77.2875}

\bibitem[{Wempe {et~al.}(2022)Wempe, Koopmans, Wierda, Hannuksela, \&
  Broeck}]{Wempe:2022zlk}
Wempe, E., Koopmans, L. V.~E., Wierda, A. R. A.~C., Hannuksela, O.~A., \&
  Broeck, C. v.~d. 2022.
\newblock \doarXiv{2204.08732}

\bibitem[{Wierda {et~al.}(2021)Wierda, Wempe, Hannuksela, Koopmans, \& Van
  Den~Broeck}]{Wierda:2021upe}
Wierda, A. R. A.~C., Wempe, E., Hannuksela, O.~A., Koopmans, L. e. V.~E., \&
  Van Den~Broeck, C. 2021, Astrophys. J., 921, 154,
  \dodoi{10.3847/1538-4357/ac1bb4}

\bibitem[{{Witt}(1990)}]{Witt:1990}
{Witt}, H.~J. 1990, \aap, 236, 311

\bibitem[{Wong {et~al.}(2021)Wong, Chan, Wong, Lo, \& Li}]{Wong:2021lxf}
Wong, H. W.~Y., Chan, L. W.~L., Wong, I. C.~F., Lo, R. K.~L., \& Li, T. G.~F.
  2021.
\newblock \doarXiv{2112.05932}

\bibitem[{Wright \& Hendry(2021)}]{Wright:2021cbn}
Wright, M., \& Hendry, M. 2021.
\newblock \doarXiv{2112.07012}

\bibitem[{Xu {et~al.}(2015)Xu, Sluse, Gao, Wang, Frenk, Mao, Schneider, \&
  Springel}]{Xu2015}
Xu, D., Sluse, D., Gao, L., {et~al.} 2015, Monthly Notices of the Royal
  Astronomical Society, 447, 3189, \dodoi{10.1093/mnras/stu2673}

\bibitem[{Xu {et~al.}(2009)Xu, Mao, Wang, Springel, Gao, White, Frenk, Jenkins,
  Li, \& Navarro}]{Xu2009}
Xu, D.~D., Mao, S., Wang, J., {et~al.} 2009, Monthly Notices of the Royal
  Astronomical Society, 398, 1235, \dodoi{10.1111/j.1365-2966.2009.15230.x}

\bibitem[{Xu {et~al.}(2021)Xu, Ezquiaga, \& Holz}]{Xu:2021bfn}
Xu, F., Ezquiaga, J.~M., \& Holz, D.~E. 2021.
\newblock \doarXiv{2105.14390}

\bibitem[{Yeung {et~al.}(2021)Yeung, Cheung, Gais, Hannuksela, \&
  Li}]{Yeung:2021roe}
Yeung, S. M.~C., Cheung, M. H.~Y., Gais, J. A.~J., Hannuksela, O.~A., \& Li, T.
  G.~F. 2021.
\newblock \doarXiv{2112.07635}

\end{thebibliography}

\appendix
\section{Conversion of the evidence between hypotheses}
\label{sec:ChangeEvidenceDerivation}
For a given hypothesis (not written explicitly here to ease the notation), the evidence for a model $M$ is
\begin{equation}\label{eq:ModelEvidence}
    \mathcal{Z}^{M} = \int d\vartheta p(D | \vartheta)p(\vartheta | M) \, ,
\end{equation}
where $D$ is the data, which can be made out of several data streams.
However, using Bayes' theorem, the posterior for a model $R$ is
\begin{equation}\label{eq:PosteriorRun}
    p(\vartheta, | D, R) = \frac{p(D | \vartheta) p(\vartheta | R)}{p(D | R)} = \frac{p(D | \vartheta) p(\vartheta | R)}{\mathcal{Z}^R} \, .
\end{equation}
So,
\begin{equation}\label{eq:pDgivenThera}
    p(D | \vartheta) = \frac{\mathcal{Z}^R p(\vartheta | D, R)}{p(\vartheta | R)} \,.
\end{equation}
Combining Eqs.~(\ref{eq:pDgivenThera}) and (\ref{eq:ModelEvidence}), one gets
\begin{align}\label{eq:DevelopChangeZ}
    \mathcal{Z}^{M} &= \int d\vartheta\mathcal{Z}^R \frac{p(\vartheta | D, R)}{p(\vartheta | R)}p(\vartheta | M) \nonumber\\
    &= \mathcal{Z}^R \int d\vartheta \frac{p(\vartheta | M)}{p(\vartheta | R)} p(\vartheta | D, M) \nonumber \\
    &= \bigg\langle \frac{p(\vartheta | M)}{p(\vartheta | R)} \bigg\rangle_{p(\vartheta | D, R)} \mathcal{Z}^R \,.
\end{align}

In practice, instead of solving the integral given in Eq.~ (\ref{eq:DevelopChangeZ}), one uses Monte Carlo integration, sampling $\vartheta$ from $p(\vartheta | D, R)$, the posterior distribution and computing a weight for each sample. The integral is then approximated as the mean of the weights:
\begin{equation}\label{eq:NumericalEvidenceConversionAppending}
    \mathcal{Z}^{M} = \frac{1}{N} \bigg( \sum_{i = 0}^{i = N} \frac{p(\vartheta^i | M)}{p(\vartheta^i | R)} \bigg) \mathcal{Z}^{R} \, ,
\end{equation}
where the $\{\vartheta_i\}_{i = 1, N}$ are sampled from $p(\vartheta | D, R)$.

\end{document}